\documentclass[a4paper,11pt]{article}
\pdfoutput=1 

\usepackage{jcappub} 
                     
\usepackage{tabularx}

\usepackage[T1]{fontenc} 

\usepackage{graphicx}
\usepackage{dcolumn}
\usepackage{bm}
\usepackage{hyperref}
\usepackage[utf8]{inputenc}
\usepackage[dvipsnames]{xcolor}
\usepackage[normalem]{ulem}
\usepackage{xspace}
\usepackage{float}
\usepackage{fontawesome}
\usepackage{multirow}
\usepackage[dvipsnames]{xcolor}
\usepackage[utf8]{inputenc}
\usepackage{blindtext}
\hypersetup{
    colorlinks=true,
    citecolor = blue,
    linkcolor=red,
    filecolor=magenta,      
    urlcolor=magenta,
}

\newcommand{\ie}{{i.e.}}

\newcommand{\eg}{{e.g.}}

\newcommand{\eq}{Eq.}

\newcommand{\fig}{Fig.}

\newcommand{\Refe}{Ref.}
\newcommand{\Refs}{Refs.}
\newcommand{\equ}[1]{\eq~(\ref{equ:#1})}
\newcommand{\figu}[1]{\fig~\ref{fig:#1}}

\title{Near-future discovery of point sources of ultra-high-energy neutrinos}

\author[a]{Damiano F.~G.~Fiorillo\note{\href{http://orcid.org/0000-0003-4927-9850}{0000-0003-4927-9850}}}
\author[a]{Mauricio Bustamante\note{\href{http://orcid.org/0000-0001-6923-0865}{0000-0001-6923-0865}}}
\author[a]{Victor B.~Valera\note{\href{http://orcid.org/ 0000-0002-0532-5766}{ 0000-0002-0532-5766}}}

\affiliation[a]{Niels Bohr International Academy, Niels Bohr Institute,\\University of Copenhagen, 2100 Copenhagen, Denmark}

\emailAdd{damiano.fiorillo@nbi.ku.dk}

\emailAdd{mbustamante@nbi.ku.dk}

\emailAdd{vvalera@nbi.ku.dk}

\abstract{Upcoming neutrino telescopes may discover ultra-high-energy (UHE) cosmic neutrinos, with energies beyond 100~PeV, in the next 10--20 years.  Finding their sources would identify guaranteed sites of interaction of UHE cosmic rays, whose origin is unknown.  We search for sources by looking for multiplets of UHE neutrinos arriving from similar directions.  Our forecasts are state-of-the-art, geared at neutrino radio-detection in IceCube-Gen2.  They account for detector energy and angular response, and for critical, but uncertain backgrounds. Sources at declination of $-45^\circ$ to $0^\circ$ will be easiest to discover.  Discovering even one steady-state source in 10~years would imply that the source has an UHE neutrino luminosity at least larger than about $10^{43}$~erg/s (depending on the source redshift evolution). Discovering no transient source would disfavor transient sources brighter than $10^{53}$~erg as dominant.  Our results aim to inform the design of upcoming detectors.}

\begin{document}
\maketitle
\flushbottom

\section{Introduction} Where do ultra-high-energy cosmic rays (UHECRs) come from?  They are the most energetic particles known---nuclei with energies in excess of $10^{12}$~GeV---yet their origin remains unknown sixty years after their discovery~\cite{Linsley:1963km,Anchordoqui:2018qom, AlvesBatista:2019tlv,Coleman:2022abf}.  As long as it does, our understanding of the high-energy Universe will be incomplete.  

\begin{figure}[h!]
    \centering
    \includegraphics[width=0.8\textwidth]{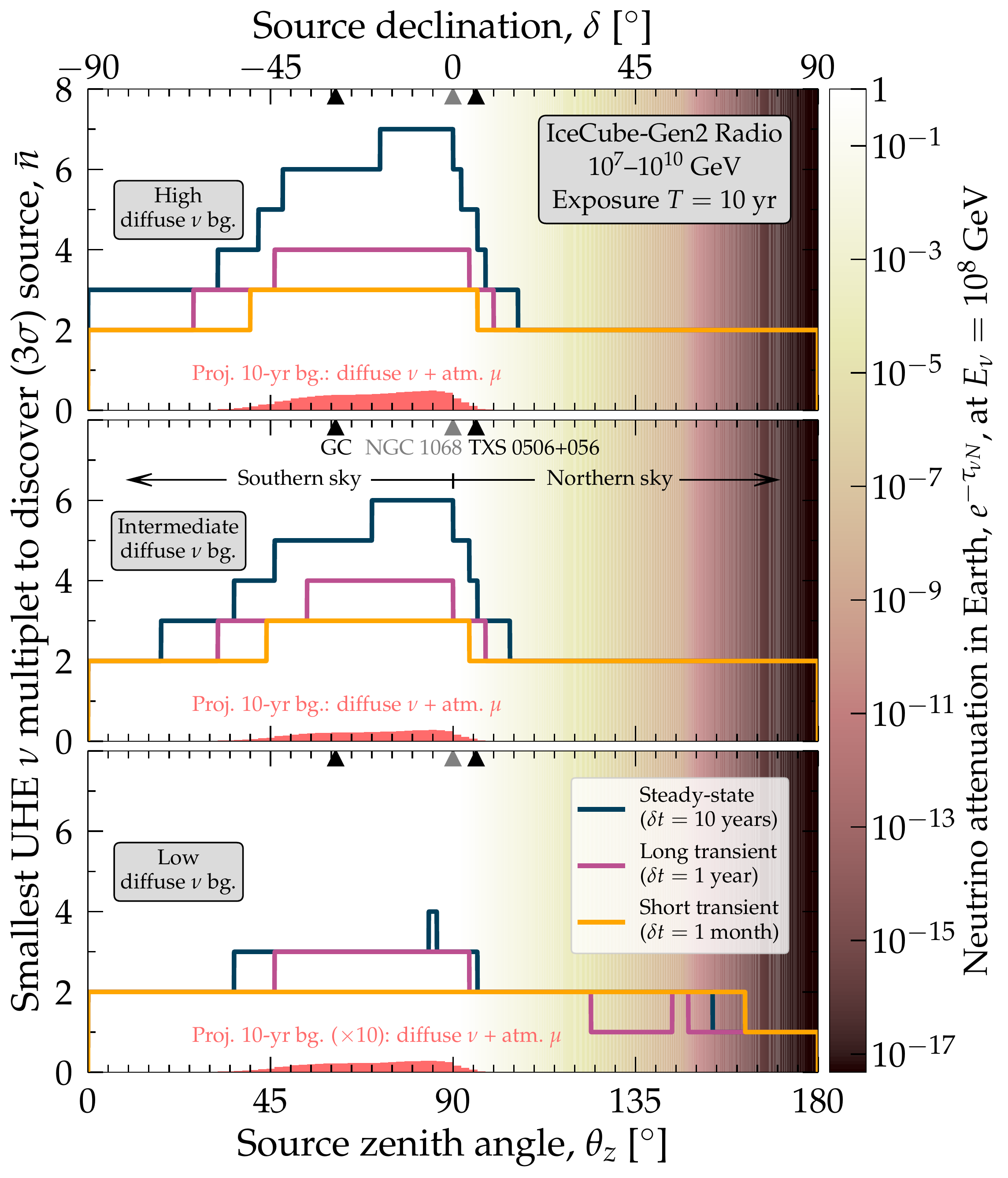}
    \caption{Smallest UHE multiplet needed for the IceCube-Gen2 radio array to discover an UHE neutrino source, steady-state or transient, with global significance of $3\sigma$, for three choices of the unknown background diffuse UHE neutrino flux: high ({\it top}), intermediate ({\it center}), and low ({\it bottom}).  For each, we show the projected 10-year rate of background events with reconstructed shower energy of $10^7$--$10^{10}$~GeV.  The shading shows the in-Earth attenuation coefficient $e^{-\tau_{\nu N}}$ for 100-PeV neutrinos, where $\tau_{\nu N}$ is the optical depth to neutrino-nucleon ($\nu N$) scattering; smaller values of it represent stronger attenuation.  For this plot, we use a detector angular resolution of $\sigma_{\theta_z} = 2^\circ$ and our baseline radio array design.  We include positions of promising sources: the Galactic Center (GC), Seyfert galaxy NGC 1068~\cite{IceCube:2021xar}, and blazar TXS 0506+056~\cite{IceCube:2018dnn}.  See the main text for details and Appendix~\ref{section:results5sigma} for $5\sigma$ results. \vspace*{-0.3cm}}
    \label{fig:histogramdetection}
\end{figure}

The sources of UHECRs are purportedly extragalactic cosmic particle accelerators located Mpc--Gpc away from Earth~\cite{PierreAuger:2017pzq,PierreAuger:2021dqp,PierreAuger:2020fbi,TelescopeArray:2014tsd,PierreAuger:2018qvk,PierreAuger:2018zqu,TelescopeArray:2020cbq,TelescopeArray:2021ygq,PierreAuger:2014yba,Kim:2021mcf,TelescopeArray:2021dfb}, though none has been irrefutably identified.  Searches for sources based on the detection of UHECRs are fundamentally limited: UHECR trajectories bend in cosmic magnetic fields~\cite{Sigl:2003ay, Dolag:2003ra, Sigl:2004yk, Dolag:2004kp, Keivani:2014kua, Hackstein:2016pwa, AlvesBatista:2017vob, Farrar:2017lhm} and, at EeV-scale energies, UHECRs rarely reach us from beyond 100~Mpc due to their scattering off cosmic photon backgrounds~\cite{Greisen:1966jv, Zatsepin:1966jv}.  
Searches based on the detection of gamma rays are similarly limited: PeV gamma rays emitted by UHECR sources get re-processed into the GeV--TeV range, where they are easily confused with backgrounds~\cite{Stecker:1971ivh, Protheroe:1996si, Coppi:1996ze, DeAngelis:2013jna}; EeV gamma rays,  more resilient, remain undiscovered~\cite{PierreAuger:2014eyz, PierreAuger:2016kuz, Fomin:2017ypo, KASCADEGrande:2017vwf, TelescopeArray:2018rbt}.

Neutrinos are free from these limitations~\cite{Ackermann:2019ows, Ackermann:2019cxh, Ackermann:2022rqc, MammenAbraham:2022xoc}.  They are made in the interaction of UHECRs with matter and radiation in the UHECR sources, and  during UHECR propagation to Earth; they receive a sizable fraction of the parent proton energy~\cite{Margolis:1977wt, Stecker:1978ah, Waxman:1998yy, Mucke:1999yb, Kelner:2006tc, Hummer:2010vx}.  Unlike UHECRs, neutrinos are electrically neutral, so they point back to their sources, and, unlike UHECRs and gamma rays, they only interact weakly, so they are not damped by interactions en route to Earth.  
Thus, detecting them may neatly reveal the positions of UHECR interaction sites in the sky.  


In the last decade, the IceCube neutrino telescope discovered high-energy cosmic neutrinos, with TeV--PeV energies~\cite{IceCube:2013cdw, IceCube:2013low, IceCube:2014stg, IceCube:2015qii, IceCube:2016umi, Ahlers:2018fkn, IceCube:2020wum}, that revealed the first likely sources of UHECRs with tens of PeV~\cite{IceCube:2018dnn, AMONTeam:2020otr, Stein:2020xhk, IceCube:2021xar}.  However, it is unknown whether these are also the long-sought sources of EeV-scale UHECRs.  To answer this definitively, we need {\it ultra-high-energy} (UHE) neutrinos, beyond 100~PeV.
Yet, they are rare, and remain undiscovered~\cite{Baltrusaitis:1984usa, Baltrusaitis:1985mt, Lehtinen:2003xv, Gorham:2003da, Abbasi:2008hr, Scholten:2009ad, James:2009sf, Kurahashi:2010ei, terVeen:2010gb, Kravchenko:2011im, PierreAuger:2012bpb, ARIANNA:2014fsk, Bray:2015lda, ARA:2015caq, MAGIC:2018gza, IceCube:2018fhm, PierreAuger:2019azx, PierreAuger:2019ens, ANITA:2019wyx, TelescopeArray:2019mzl, ARA:2019wcf, PierreAuger:2020llu, Ogawa:2021tdn, Krampah:2021ysn} since their prediction in the 1960s~\cite{Greisen:1966jv, Zatsepin:1966jv}.  

Fortunately, upcoming UHE neutrino telescopes have a real chance of discovering them in the next 10--20 years, even if their flux is low~\cite{Ackermann:2019ows, MammenAbraham:2022xoc, Ackermann:2022rqc, Valera:2022wmu}.  We capitalize on this by making state-of-the-art forecasts of the UHE source discovery via UHE {\it neutrino multiplets}, \ie, clusters of neutrinos from similar positions in the sky, indicative of a source.  We gear our forecasts to IceCube-Gen2~\cite{IceCube-Gen2:2020qha}, one of the leading upcoming neutrino telescopes, expected to start operations in the 2030s. 

We provide methods and baseline forecasts to inform the aims and designs of upcoming UHE neutrino telescopes.  We frame our results in terms of two questions: what is the size of the smallest UHE multiplet needed to claim source discovery, and what would source discovery, or lack thereof, imply for the population of UHE sources.

Figure~\ref{fig:histogramdetection} shows our answer to the first question.  The variation with source position reflects the angular distribution of background events, which itself reflects the in-Earth neutrino attenuation and detector response.  Results depend strongly on the size of the diffuse UHE neutrino flux, presently unknown, which is the main background that the source search must overcome: a higher background demands larger multiplets.  Later, in \figu{pointsourcesensitivity}, we show our answer to the second question above.


Pioneering work explored the prospects of discovering point sources of high-energy~\cite{Lipari:2008zf, Silvestri:2009xb, Murase:2012df, Ahlers:2014ioa, Murase:2016gly, Fang:2016hyv, Bartos:2021tok} and UHE neutrinos~\cite{Fang:2016hop}.  Our methods add key features that enhance its usefulness to realistic UHE source searches.  Individually, they lead to sizeable improvements; together, they lead to powerful advances. 

First, we avoid introducing source-model bias by making the diffuse UHE neutrino flux equal to different plausible choices, rather than modeling it as coming from the same source population responsible for the multiplets.  Second, we account for the atmospheric muon background, recently found to matter for UHE neutrino radio-detection~\cite{Garcia-Fernandez:2020dhb}.  Third, we account for critical, but often-overlooked features in source searches, \eg, neutrino propagation through Earth, the energy- and direction-dependent detector response, and its energy and angular resolution~\cite{Valera:2022ylt}.  Fourth, though our methods apply generally, we ground our forecasts in the radio-detection of neutrinos in IceCube-Gen2, via state-of-the-art simulations of neutrino interactions, the ensuing particle showers, and the emission, propagation, and detection of radio signals~\cite{Glaser:2019cws}.

\Refs~\cite{Fang:2016hyv, Fang:2016hop} use the sky-wide distribution of the angular separation of neutrino pairs to detect the presence of point sources; \Refe~\cite{Fang:2016hyv} used them also to locate sources, though not UHE ones.  Unlike them, we tessellate the sky to rely only on local information and use multiplets---not only pairs---{\it mainly} to locate point sources.  Appendix~\ref{section:appendixprobabilities} extends our methods to resemble \Refe~\cite{Fang:2016hop}. Our approach is analytic, complementary to the Monte Carlo approach used by \Refe~\cite{Palladino:2019hsk} for TeV--PeV neutrinos.


\section{Detecting UHE neutrinos}  We compute the propagation of UHE neutrinos inside Earth and their detection in IceCube-Gen2 as in \Refe~\cite{Valera:2022ylt}.  Below, we sketch the methods; for details, see Appendix~\ref{section:appendixdetectionrate} and \Refe~\cite{Valera:2022ylt}.

We model UHE neutrino propagation through the Earth using the state-of-the-art code {\sc NuPropEarth}~\cite{Garcia:2020jwr, NuPropEarth}. The detector response is modeled via its effective volume, Fig.~13 in \Refe~\cite{Valera:2022ylt}, simulated using the same tools, {\sc NuRadioMC}~\cite{Glaser:2019cws} and {\sc NuRadioReco}~\cite{Glaser:2019rxw}, as the IceCube-Gen2 Collaboration. For our main results, we adopt the baseline design of the IceCube-Gen2 radio array from \Refe~\cite{IceCube-Gen2:2021rkf}: 313 detector stations, made up of 169 shallow stations and 144 hybrid, \ie, shallow plus deep, stations with complementary response.  Appendix~\ref{section:experimentalconfiguration} contains results for alternative designs; our conclusions are broadly unaffected. Our choices of detector resolution are informed by simulations~\cite{Anker:2019zcx, Glaser:2019rxw, Gaswint:2021smu, RNO-G:2021zfm, Stjarnholm:2021xpj, ARIANNA:2021pzm, Aguilar:2021uzt, Gaswint:2021smu, Stjarnholm:2021xpj, ARA:2021bss, Valera:2022ylt}; they are the same as in \Refe~\cite{Valera:2022ylt}.


\section{Backgrounds}  The main challenge to multiplet searches is that, underlying the UHE neutrinos from point sources, we expect a diffuse background of UHE neutrinos and atmospheric muons whose random over-fluctuations may mimic multiplets from point sources.  Later, we show how our methods overcome this.  

The diffuse flux of UHE neutrinos is likely composed of cosmogenic neutrinos~\cite{Greisen:1966jv, Zatsepin:1966jv, Berezinsky:1969erk}, made in UHECR interactions en route to Earth, and of neutrinos from unresolved sources.  There are numerous flux predictions for both~\cite{Fang:2013vla, Padovani:2015mba, Fang:2017zjf, Romero-Wolf:2017xqe, AlvesBatista:2018zui, Heinze:2019jou, Muzio:2019leu, Rodrigues:2020pli, Anker:2020lre, Muzio:2021zud}, and upper limits~\cite{Baltrusaitis:1984usa, Baltrusaitis:1985mt, Lehtinen:2003xv, Gorham:2003da, Abbasi:2008hr, Scholten:2009ad, James:2009sf, Kurahashi:2010ei, terVeen:2010gb, Kravchenko:2011im, PierreAuger:2012bpb, ARIANNA:2014fsk, Bray:2015lda, ARA:2015caq, MAGIC:2018gza, IceCube:2018fhm, PierreAuger:2019azx, PierreAuger:2019ens, ANITA:2019wyx, TelescopeArray:2019mzl, ARA:2019wcf, PierreAuger:2020llu, Ogawa:2021tdn, Krampah:2021ysn} come chiefly from IceCube~\cite{IceCube:2018fhm} and Auger~\cite{PierreAuger:2019ens}.  Rather than adopting a particular prediction, we set the diffuse UHE neutrino flux to benchmark levels representative of current and future detector sensitivity: the current IceCube upper limit on the energy flux $E_\nu^2 \Phi_\nu$ ({\it high})~\cite{IceCube:2018fhm}, and versions of it shifted down to $10^{-8}$ ({\it intermediate}) and $10^{-9}$~GeV~cm$^{-2}$~s$^{-1}$~sr$^{-1}$ ({\it low}).  Each represents a future possibility for the largest allowed diffuse UHE neutrino flux, either a flux measurement or an upper limit. We conservatively do not perform a spectral analysis in energy.   The above benchmark fluxes yield roughly 520, 297, and 30 events all-sky, respectively, in 10 years of IceCube-Gen2; see Appendix~\ref{section:detailsonbackground}.

The diffuse neutrino flux is isotropic, but the angular distribution of the event rate is not.  Fewer events come from directly above the detector ($\theta_z \lesssim 45^\circ$), due to weaker detector response, and from below the detector ($\theta_z \gtrsim 95^\circ$), due to in-Earth neutrino attenuation, than from slant and horizontal directions ($45^\circ \lesssim \theta_z \lesssim 95^\circ$). We also account for a background of atmospheric muons, which is however negligible; see Appendix~\ref{section:detailsonbackground}.  Because of the anisotropic event rate, source discovery prospects vary across the sky.


\section{Discovering sources}  The first question that we address is how large should a detected multiplet be to claim that it is due to a point source, and not to an over-fluctuation of the background.

We tessellate the sky into $N_{\rm pixels}$ square pixels that approximate the circular error regions expected from shallow detector stations~\cite{IceCube-Gen2:2021rkf}.  The width in zenith angle of each pixel is $\sigma_{\theta_z}$, the detector angular resolution. The width in azimuth is chosen so that the solid angle of each pixel equals that of a cone of apex angle $2\sigma_{\theta_z}$.  In the $i$-th pixel, we compute the mean  number of background-induced events, $\mu_i$, expected after an exposure time $T$, following the procedure sketched above. 

The {\it local} p-value $p$ of detecting a multiplet of more than $n_i$ events in the $i$-th pixel, \ie, the probability that a multiplet is due to background alone, is
$p(\mu_i,n_i)=\sum_{k=n_i}^{+\infty}  (\mu_i^k/k!)e^{-\mu_i}$.  But this does not account for the look-elsewhere effect: even if $p$ is small---so that a background fluctuation is unlikely---the probability that an excess with this p-value occurs anywhere in the sky may be large.  Therefore, in our forecasts we use instead the {\it global} p-value $P(p)$, \ie, the probability that a multiplet with local p-value $p$ occurs in any of the pixels, \ie,
\begin{equation}
 \label{equ:globalpvaluedef}
 P(p)
 =
 1 -
 \prod_{i=1}^{N_{\rm pixels}}
 \left(1-\sum_{k=\bar{n}_i(p)}^{\infty} \frac{\mu_i^k}{k!}e^{-\mu_i}\right).
\end{equation}
Appendix~\ref{section:appendixprobabilities} contains the derivation of \equ{globalpvaluedef}.  Roughly, $P \sim N_{\text{pixels}} p$, provided $p \ll N_{\text{pixels}}^{-1}$.  Given a target global p-value $\bar{P}$, we invert \equ{globalpvaluedef} to find the  local p-value $\bar{p}$ and the size of the smallest multiplet in each pixel needed to reach it, $\bar{n}_i(\bar{p})$.  We report this in \figu{histogramdetection} and in Appendix~\ref{section:appendixprobabilities}. For transient sources, with an emission period of duration $\delta t \ll T$, \eg, a flaring blazar or gamma-ray burst, we modify this procedure to account for a look-elsewhere effect in time; see Appendix~\ref{section:appendixprobabilities}.

Figure~\ref{fig:histogramdetection} shows the smallest multiplet needed to claim source discovery at $3 \sigma$, \ie, with $\bar{P} = 0.003$, in $T = 10$~yr of exposure time, for our three background neutrino diffuse fluxes.  In all cases, sources located above IceCube-Gen2 ($\theta_z \lesssim 45^\circ$), where the  background is smallest, may be discovered by detecting a doublet or triplet, regardless of the choice of benchmark.  However, detection is unlikely in these directions because the detector effective volume is small.  In contrast, sources located closer to the horizon ($45^\circ \lesssim \theta_z \lesssim 95^\circ$), where the background is largest, require larger multiplets, as large as a heptaplet for the high background benchmark.  Yet, detection is promising in these directions because the effective volume is larger and in-Earth attenuation is mild.

Steady-state sources, like starburst galaxies, are active during the full exposure time, $T$.  Searches for them accumulate larger background and require larger multiplets to claim discovery.  Long- and short-duration transient sources, like blazar flares and tidal disruption events, respectively, are active for a fraction of that time, $\delta t$.  Searches for them require smaller multiplets.  For very-short-duration transients, like gamma-ray bursts, doublets or triplets are always enough, regardless of source position and background level; see Appendix~\ref{section:sourcepopulationvaryingbackground}.

{\it Our results are significantly affected by the angular resolution of the detector.}  Figure~\ref{fig:histogramdetection} uses $\sigma_{\theta_z} = 2^\circ$.  Better angular resolution allows for finer sky pixels and a smaller contribution of the background in each of them.  Using $\sigma_{\theta_z} = 5^\circ$ roughly doubles the size of the multiplets needed to claim discovery; see Appendix~\ref{section:experimentalconfiguration}.

Our results are tentatively robust to the choice of the design of the IceCube-Gen2 radio array. Figure~\ref{fig:histogramdetection} uses our baseline design described above.  At least for two alternative designs, our general observations hold; see Appendix~\ref{section:experimentalconfiguration}.  Our claim stems from a non-exhaustive exploration of array designs.  Stronger claims require further simulation work, which is ongoing~\cite{IceCube-Gen2:2021rkf}.


\section{Multiplets from a population of UHE sources}
\label{section:appendixexpectedmultiplets}

Next we show what we can learn about the UHE neutrino source populations with the detection or absence of a multiplet. We compute how likely it is to discover a source from a given source population.  Following what seminal \Refs~\cite{Silvestri:2009xb, Murase:2016gly} did for TeV--PeV neutrino sources, we consider a population of identical UHE sources distributed in redshift, $z$.  For steady-state sources, we describe the population using the neutrino luminosity emitted by a source, $L_\nu$, and the local source number density, $n_0 \equiv n(z=0)$. The parameter $L_\nu$, in particular, significantly depends on the internal parameters of the source and on the chemical composition of the UHECRs.  For transient sources, we use the energy emitted by a source in neutrinos, $E_\nu$, and the local source burst rate, $\mathcal{R}_0 \equiv \mathcal{R}(z=0)$.  All sources in a population share the same value of $L_\nu$ or $E_\nu$.  Each source emits neutrinos with a broken power-law spectrum that approximates neutrino production via proton-photon interactions~\cite{Dermer:2012rg, Winter:2012xq, Murase:2014foa, Fiorillo:2021hty}.  The neutrino emissivity, $L_\nu n(z)$ or $E_\nu \mathcal{R}(z)$, follows the star-formation rate~\cite{Hopkins:2006bw, Murase:2016gly}.  Fixing the values of the source population parameters fixes the neutrino flux coming from that source class. 

Let us denote the number density per comoving volume as $n(z) = n_0 f(z)$, where $n_0 \equiv n(z=0)$ is the local number density and $f(z)$ describes the evolution with redshift $z$.  We set $f(z)$ to be equal to the star-formation rate~\cite{Hopkins:2006bw,Murase:2016gly}, normalized so that $f(z=0) = 1$. (For specific source classes, the redshift evolution of the sources might differ, \eg, for FSRQs~\cite{Ajello:2011zi} and BL Lacs~\cite{Ajello:2013lka}.  We examine how this impacts our results in Appendix~\ref{section:effectofevolution}.)  For simplicity, we assume that all the  sources in a given population have the same broken-power-law neutrino spectrum,
\begin{equation}
 \frac{dN_\nu}{dEdt}
 \propto
 \left[
 a \left(\frac{E}{E_0}\right)^\alpha+(1-a) 
 \left(\frac{E}{E_0}\right)^\beta
 \right]^{-1} \;,
\end{equation}
normalized so that $\int_0^\infty dE E (dN_\nu/dEdt) = L_\nu$. For transient sources, the equivalent parametrization is in terms of the total number of neutrinos injected in the burst, $dN_\nu/dE$, normalized to the total energy of the burst, \ie, $\int_0^\infty dE E (dN_\nu/dE)=E_\nu$  To produce our results, we choose $\alpha=1$, $\beta=3$, $a=0.8$, and $E_0=10^8$~GeV, which approximates a neutrino energy spectrum that peaks inside the energy range where the IceCube-Gen2 radio array is sensitive.  Inside one pixel, the probability distribution of sources in redshift is
\begin{equation}
 \label{equ:redshiftdistribution}
 p_{\rm src}(z)
 =
 \frac{\delta\Omega ~ n_0 ~ f(z) r^2(z)}{H(z) N_{\mathrm{src}}} \;,
\end{equation}
where $\delta\Omega$ is the solid angle of a pixel, $r(z)$ is the comoving distance, and $H(z)$ is the Hubble parameter.  We assume a $\Lambda$CDM cosmology, with Hubble constant $H_0 = 67.4$~km~s$^{-1}$~Mpc$^{-1}$, and adimensional energy density parameters $\Omega_m = 0.315$, $\Omega_\Lambda = 0.685$~\cite{ParticleDataGroup:2020ssz}. Finally, the mean number of sources in each pixel is 
\begin{equation}
 N_{\rm src} 
 \label{equ:number_sources_per_pixel}
 = 
 \int_0^\infty dz ~ \frac{n_0~f(z)~\delta\Omega~r^2(z)}{H(z)} \;.
\end{equation}

Below, we detail how we compute the constraints on the source population parameters that we show in the main text.  The procedure consists of three steps: computing the contribution of the diffuse background in each pixel, computing the distribution of the number of events in each pixel, and computing constraints on the source populations.

\let\oldaddcontentsline\addcontentsline
\renewcommand{\addcontentsline}[3]{}
\subsection{Contribution of the diffuse background in each pixel}
\let\addcontentsline\oldaddcontentsline

The normalization and angular distribution of the neutrino-induced event rate from a source depends on the source redshift, source luminosity, and zenith angle of the pixel in which the source is located.  Formally, it also depends on the neutrino energy spectrum, since different neutrino energy spectra are affected differently by their propagation through the Earth.  However, because our analysis groups events in a single energy bin, it is largely insensitive to the shape of the neutrino energy spectrum.  
A source at redshift $z$ produces a mean number of events $s_i(z)$ in the $i$-th pixel. The number of events from this source in different pixels is different because of different effects of in-Earth attenuation and because of the angular response of the detector.  We account for these differences by assuming that $s_i(z) = \kappa_s B_i$, where $B_i$ is the mean number of events expected from a particular choice of the diffuse isotropic UHE neutrino background (see Appendix~\ref{section:detailsonbackground}), and $\kappa_s$ is a proportionality constant that we determine below.  We compute the proportionality constant between $s_i$ and $B_i$ using the following procedure:
\begin{itemize}
 \item
  First, we compute $B_i$ in each pixel, following our sophisticated procedure from Appendix~\ref{section:appendixdetectionrate}.  We pick either of our benchmark background isotropic UHE neutrino fluxes---low, intermediate, or high; see \figu{diffusebackground}.  All of them yield the same angular distribution of events; they differ only in the normalization of the event rate; 
 \item
  From the projected 90\% C.L.~sensitivity to the diffuse UHE neutrino flux of the radio array of IceCube-Gen2 reported in \Refe~\cite{IceCube-Gen2:2021rkf}, we extract the energy-dependent effective area, $A'_{\text{eff}}(E)$, following Eq.~(C.1) in \Refe~\cite{Fiorillo:2021hty}.  The treatment in \Refe~\cite{Fiorillo:2021hty} is geared to the sensitivity to point sources, expressed in GeV~cm$^{-2}$~s$^{-1}$.  To extract the effective area from the sensitivity to the diffuse flux, measured in GeV~cm$^{-2}$~s$^{-1}$~sr$^{-1}$, we include an additional factor of $4\pi$ in the denominator of Eq.~(C.1) in \Refe~\cite{Fiorillo:2021hty};
 \item 
  Because the effective area extracted above corresponds to a different detector array design than the ones we use, the event rate computed using it does not match the event rate computed following our procedure from Appendix~\ref{section:appendixdetectionrate}.  To fix this, we scale the effective area by an energy-independent factor $\kappa_A$, defined implicitly as $A_{\rm eff}(E) = \kappa_A A^\prime_{\rm eff}(E)$, and chosen so that the all-sky rate computed using $A_{\rm eff}$ for our chosen background model (see below) matches  $\sum_{i=1}^{N_{\text{pixels}}} B_i$.  (We use this workaround only when computing constraints on the source population.  When computing the smallest multiplets needed to claim source discovery, we use directly our procedure from Appendix~\ref{section:appendixdetectionrate});
 \item
  The effective area that we extracted from \Refe~\cite{IceCube-Gen2:2021rkf} is for a diffuse isotropic neutrino flux; the angular response of the detector has already been averaged over all the sky.  Therefore, with this effective area we compute the mean number of events from a source at redshift $z$ when the source direction is averaged over all the sky.  For a steady-state source, this is
  \begin{equation}
   \label{equ:averagenumber}
   \langle s(z) \rangle
   =
   \frac{T}{4\pi r^2(z)}
   \int_0^{\infty}
   \frac{dN_\nu}{dEdt}[E(1+z)] 
   A_{\text{eff}}(E) dE \;,
  \end{equation}
  where $T=10$~years is the exposure time.  For a transient source, it is
  \begin{equation}
   \label{equ:averagenumbertransient}
   \langle s(z) \rangle
   =
   \frac{1}{4\pi r^2(z)}
   \int_0^{\infty}
   \frac{dN_\nu}{dE}[E(1+z)] 
   A_{\text{eff}}(E) dE \;;
  \end{equation}
 \item
  Finally, we compute $\kappa_s$, the proportionality constant between $s_i(z)$ and $B_i$ by requiring that the average of $s_i(z)$ over all pixels, $\kappa_s \sum_{j=1}^{N_{\text{pixels}}} B_j / N_{\rm pixels}$, matches \equ{averagenumber} or \equ{averagenumbertransient}, which yields
  \begin{equation}
   s_i(z)
   =
   \frac{\langle s(z) \rangle N_{\text{pixels}}}{\sum_{j=1}^{N_{\text{pixels}}} B_j} B_i \;.
  \end{equation}
\end{itemize}
The mean number of events in the $i$-th pixel is
\begin{equation}
 \label{equ:mean_number_events_pixel_i}
 \langle n_i \rangle
 =
 N_{\rm src} \int_0^{\infty} dz~p_{\rm src}(z) s_i(z)
 +
 b_i \;,
\end{equation}
where the first term is due to point sources, calculated following the above procedure and the second term, $b_i$, is due to the background of cosmogenic neutrinos, unresolved sources, and atmospheric muons.  The value of $b_i$ is fixed by requiring that the mean event rate per pixel is saturated by the background, \ie, $\langle n_i \rangle = B_i$.

\let\oldaddcontentsline\addcontentsline
\renewcommand{\addcontentsline}[3]{}
\subsection{Computing the distribution of the  number of events in each pixel}
\let\addcontentsline\oldaddcontentsline

Next, we compute the probability distribution of the number of events $n_i$, detected, in the $i$-th pixel.  This is given by the convolution of the Poisson distribution of the number of sources, $\sigma_i$, centered around the mean number of sources per pixel, $N_{\rm src}$, from \equ{number_sources_per_pixel}; the probability density, $p_{\rm src}(z_\alpha)$, from \equ{redshiftdistribution}, 
for the source number $\alpha$ in this pixel, with redshift $z_\alpha$, where $\alpha=1, \ldots, \sigma_i$; the Poisson distribution of the number of detected events, $n_i$, centered around the mean value $b_i + \sum_{\alpha=1}^{\sigma_i} s_i(z_\alpha)$.  Altogether, the probability distribution is
\begin{equation}
 P_i(n_i)
 =
 \sum_{\sigma_i=0}^{\infty} 
 \frac{N_{\rm src}^{\sigma_i}e^{-N_{\rm src}}}{\sigma_i!}
 \prod_{\alpha=1}^{\sigma_i} 
 \int_0^{\infty} p_{\rm src}(z_\alpha) dz_\alpha 
 \frac{(b_i+\sum_{\alpha=1}^{\sigma_i} s_i(z_\alpha))^{n_i}}{n_i!} 
 e^{-b_i-\sum_{\alpha=1}^{\sigma_i} s_i(z_\alpha)} \;.
\end{equation}
To simplify numerical evaluation, we find an analytical expression for $P_i(n_i)$ by computing its generating function, $\Phi_i(x)=\sum_{n_i=0}^{\infty} P_i(n_i) x^{n_i}$.  The sum over $n_i$ can then be performed explicitly, \ie,
\begin{equation}
 \Phi_i(x)
 =
 \exp\left[
 b_i (x-1)
 +
 N_{\rm src}
 \left(
 \int_0^{\infty} dz~p_{\rm src}(z) e^{s_i(z)(x-1)} -1
 \right)
 \right] \;.
\end{equation}
Thus, by the definition of the generating function, the probability distribution is
\begin{equation}
 \label{equ:prob_analytical}
 P_i(n_i)
 =
 \frac{1}{n_i!}
 \left[\frac{d^{n_i} \Phi_i(x)}{dx^{n_i}}
 \right]_{x=1} \;.
\end{equation}
Below, when finding constraints on the source population, we use \equ{prob_analytical} to compute numerically the probability of detecting $n_i$ events in the $i$-th pixel.

\let\oldaddcontentsline\addcontentsline
\renewcommand{\addcontentsline}[3]{}
\subsection{Computing constraints on the source populations}
\let\addcontentsline\oldaddcontentsline

Finally, we set constraints on the source population parameters, \ie, on $n_0$ and $L_\nu$ for steady-state sources, or $\mathcal{R}_0$ and $E_\nu$ for transient sources; see the main text.  When setting constraints on the source population, our underlying assumption is of source discovery with a global significance of $5\sigma$.  Requiring this fixes the size of the smallest multiplet needed to claim source discovery in each pixel, $n_{i, 5\sigma}$; see Fig.~1 in the main text.  Therefore, the probability that no source is discovered in the $i$-th pixel is
\begin{equation}
 Q_i
 =
 \sum_{n_i=0}^{n_i=n_{i, 5\sigma}-1} P_i(n_i) \;.
\end{equation}
\begin{figure}[t]
    \centering
    \includegraphics[width=0.5\textwidth]{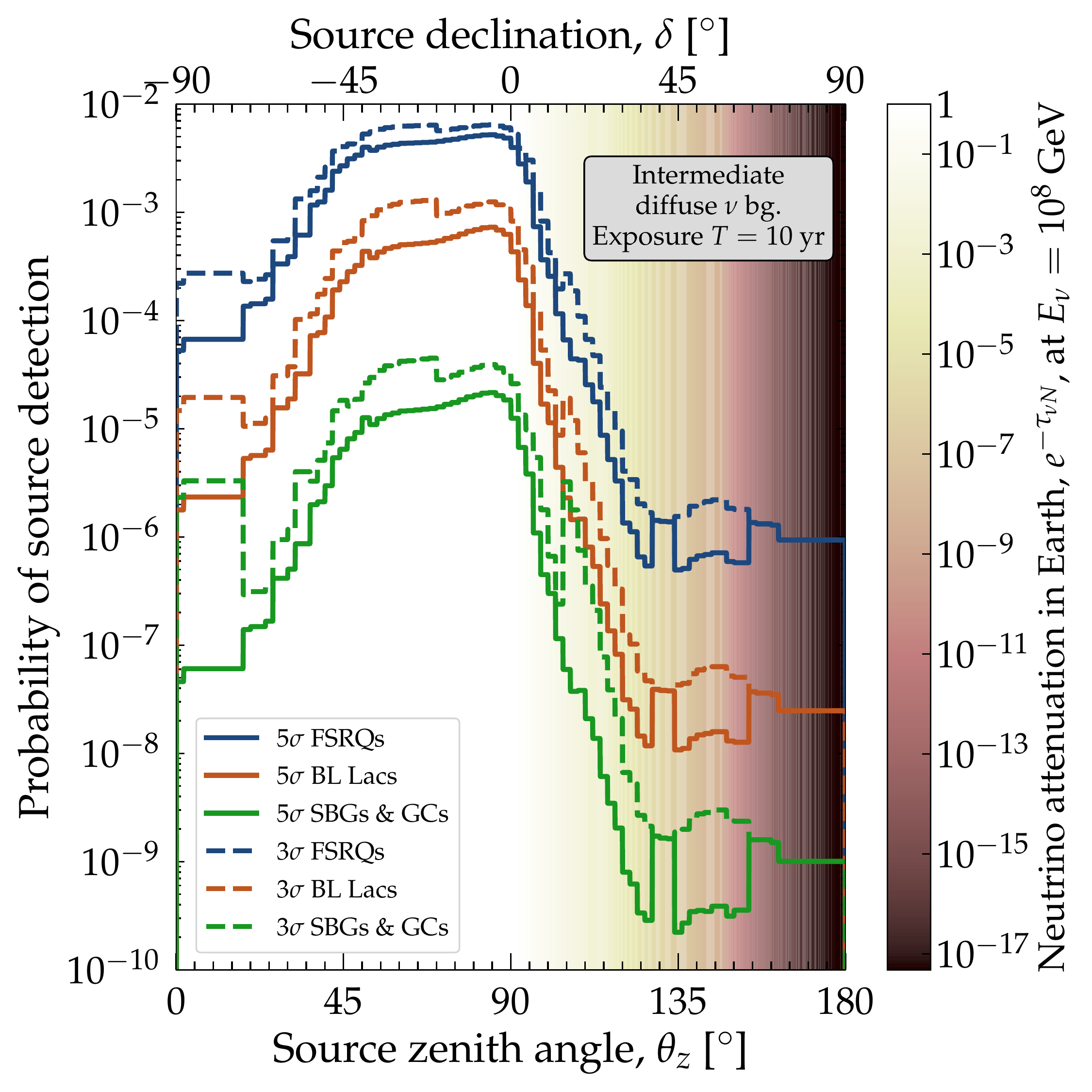}
    \caption{Probability of source discovery in each pixel in the sky, at $3\sigma$ and $5\sigma$, for three candidate classes of steady-state UHE neutrino sources: FSRQs, BL Lacs, and SBGs \& GCs.  For this plot, the detector angular resolution is $\sigma_{\theta_z} = 2^\circ$, and we use our baseline array design and intermediate benchmark diffuse neutrino flux. The large jumps at high zenith angles are due to statistical fluctuations in the multiplet size needed for detection.}
    \label{fig:detectionprobability}
\end{figure}

With this, the probability that no source is discovered anywhere in the sky is simply the product of $Q_i$ over all pixels, \ie,
\begin{equation}
 \label{equ:prob_no_source_global}
 \mathcal{Q}
 =
 \prod_{i=1}^{N_{\text{pixels}}} Q_i \;.
\end{equation}

We now use this probability to identify the regions of parameter space which are expected to lead to the discovery of point sources.

\subsection{Results}

\begin{figure}[H]
 \centering
 \includegraphics[scale=0.4]{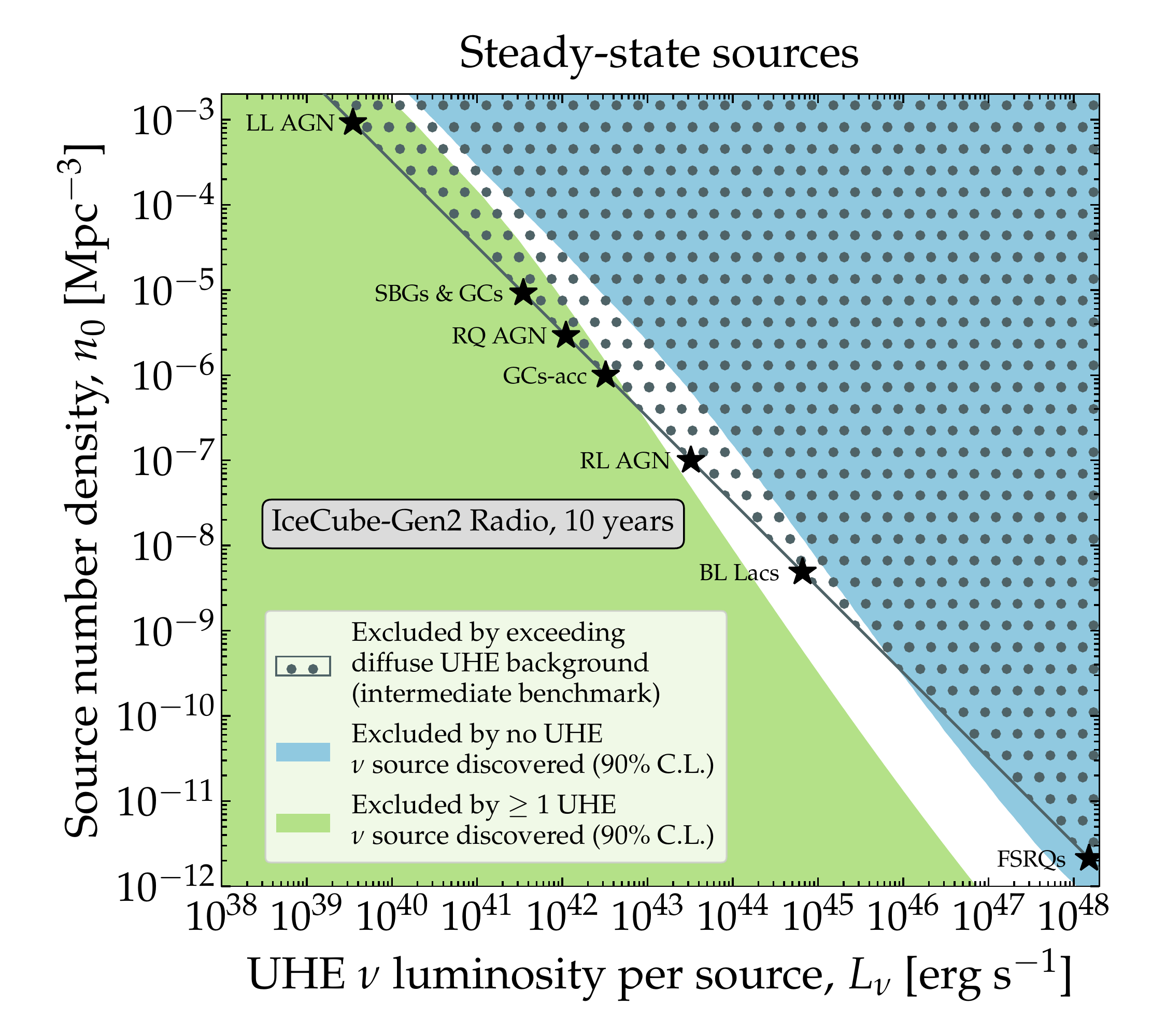}
 {\vspace{0.1cm}}
 \includegraphics[scale=0.4]{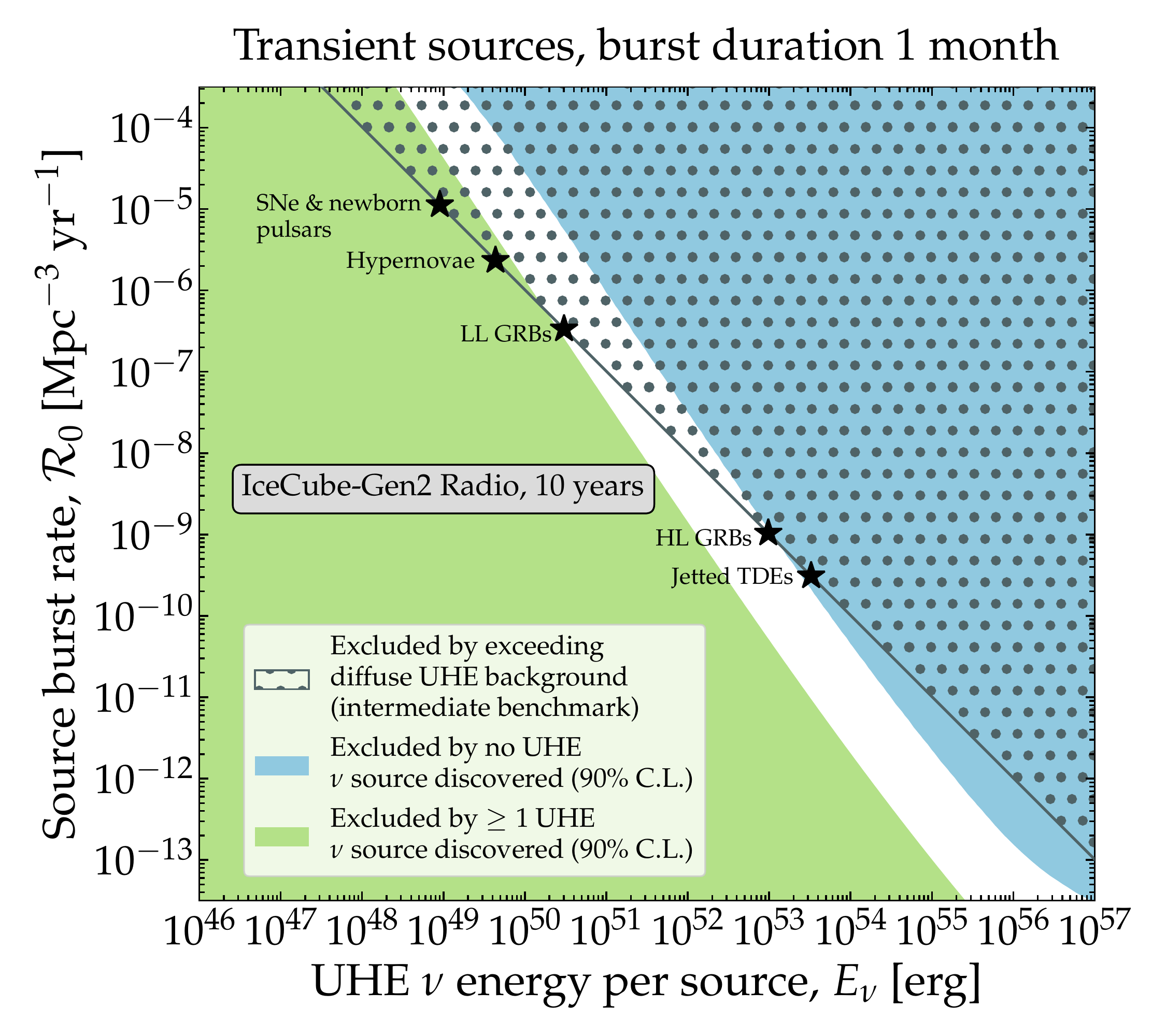}
 \caption{Constraints on candidate classes of steady-state ({\it top}) and transient ({\it bottom}) UHE neutrino sources, from the discovery or absence of UHE multiplets, \ie, of UHE neutrino sources, in the radio array of IceCube-Gen2 after 10 years of exposure.  We showcase promising candidate source classes: low-luminosity active galactic nuclei~\cite{Kimura:2014jba,Kimura:2020thg} (LL AGN), starburst galaxies~\cite{Loeb:2006tw,Tamborra:2014xia,Palladino:2018bqf,Peretti:2018tmo,Peretti:2019vsj,Ambrosone:2020evo} \& galaxy clusters~\cite{Murase:2008yt,Kotera:2009ms,Berezinsky:1996wx} (SBGs \& GCs), radio-quiet AGN~\cite{Alvarez-Muniz:2004xlu} (RQ AGN), accretion shocks in galaxy clusters~\cite{Keshet:2002sw,Kushnir:2009vm,Fang:2016amf} (GCs-acc), radio-loud AGN~\cite{Mannheim:1995mm,Murase:2014foa,Neronov:2020fww} (RL AGN), BL Lacertae AGN~\cite{Palladino:2018lov,Rodrigues:2020pli} (BL Lacs), flat-spectrum radio quasars~\cite{Atoyan:2001ey,Atoyan:2002gu,Palladino:2018lov,Righi:2020ufi} (FSRQs), supernovae~\cite{Drury:1993pd,Alvarez-Muniz:2002con,Villante:2008qg,Xiao:2016rvd,Murase:2017pfe,Tamborra:2018upn} \& newborn pulsars~\cite{Fang:2013vla,Fang:2014qva} (SNe \& newborn pulsars), hypernovae~\cite{Wang:2007ya,Xiao:2016rvd}, low-luminosity gamma-ray bursts~\cite{Murase:2006mm,Senno:2015tsn} (LL GRBs), high-luminosity GRB~\cite{Paczynski:1994uv,Waxman:1997ti,Bustamante:2014oka,Pitik:2021xhb,Guarini:2021gwh} (HL GRBs), and jetted tidal disruption events~\cite{Farrar:2008ex,Wang:2011ip,Dai:2016gtz,Senno:2016bso,Lunardini:2016xwi,Zhang:2017hom,Guepin:2017abw,Winter:2020ptf} (jetted TDEs). For each, the value of $n_0$ or $\mathcal{R}_0$ is from \Refs~\cite{Murase:2016gly, Ackermann:2019ows}; the value of $L_\nu$ or $E_\nu$ is chosen to saturate the background UHE neutrino diffuse flux; for this plot, we fix it to our intermediate benchmark.  Appendix~\ref{section:sourcepopulationvaryingbackground} contains results using our low and high benchmarks.  In the hatched region, the flux from the source population exceeds the background diffuse flux.  See the Appendix~\ref{section:appendixexpectedmultiplets} for the values of the source-population parameters. } 
 \label{fig:pointsourcesensitivity}
\end{figure}

Figure~\ref{fig:pointsourcesensitivity} shows the constraints that we set on UHE source populations based on \equ{prob_no_source_global}, assuming our intermediate benchmark UHE diffuse background. 

The first constraint comes from demanding that the neutrino flux from each class saturates the background diffuse neutrino flux. Doing this fixes the values of $L_\nu$ and $E_\nu$; the values of $n_0$ and $\mathcal{R}_0$ are from \Refe~\cite{Murase:2016gly, Ackermann:2019ows}.  Tables~\ref{tab:steady} and~\ref{tab:transient} shows the values of the population parameters of each source class for our three benchmark background fluxes. In \Refe~\cite{Murase:2016gly}, the values of $n_0$ and $\mathcal{R}_0$ were obtained under the assumption of a model-dependent scaling between the high-energy neutrino and gamma-ray luminosity, which we do not apply (however, see Appendix~\ref{section:effectofevolution} for a similar treatment of FSRQs as that of \Refe~\cite{Murase:2016gly}).  Further, for some of the source classes that we show, UHE neutrino production may not even be expected at all.  We forego a discussion of the different models of UHE neutrino production in the different candidate source classes, which is not our goal.  Since all the candidate source classes in \figu{pointsourcesensitivity} are treated on equal footing, without modeling the physical conditions and neutrino production mechanisms particular to each, the values of the population parameters in it should be taken merely as indicative.

The second constraint comes from demanding that each class either yields at least one $5\sigma$ source discovery, or none, with $90\%$ probability. The blue region would be excluded by the lack of $5\sigma$ source discovery, and is obtained by demanding 
$\mathcal{Q} < 0.1$.  The second region would be excluded by the discovery, at $5\sigma$, of at least one source, and is obtained by demanding 
$\mathcal{Q} > 0.9$.  Figure~\ref{fig:pointsourcesensitivityvaryingbackground}, in Appendix \ref{section:sourcepopulationvaryingbackground}, shows equivalent results for our low and high benchmarks.

Versions of \figu{pointsourcesensitivity} for TeV--PeV sources were shown, \eg, in \Refs~\cite{Silvestri:2009xb, Murase:2016gly, Ackermann:2019ows}; to our knowledge, \figu{pointsourcesensitivity} shows for the first time multiplet-based constraints for UHE sources.

For steady-state sources, if none is discovered after 10 years, only bright, rare source classes would be disfavored as individually dominant. While \figu{pointsourcesensitivity} may suggest that FSRQs would be excluded, in reality some candidate source classes, like TDEs, FSRQs and BL Lacs, have a redshift evolution that is quite different from the star-formation rate.  Appendix~\ref{section:effectofevolution} illustrates this.  Conversely, if even one source is discovered, most known candidate source classes would be excluded.

\begin{table}[b!]
    \centering
    \begin{tabular}{ccccc}
       Source class  & $n_0$~[Mpc$^{-3}$] & $L_\nu$~[erg~s$^{-1}$] & $L_\nu$~[erg~s$^{-1}$] & $L_\nu$~[erg~s$^{-1}$] \\
        &  & (low $\nu$ bg.) & (interm. $\nu$ bg.) & (high $\nu$ bg.)\\
       \hline
       LL AGN & $9.2\times 10^{-4}$ & $3.5\times 10^{38}$ & $3.5\times10^{39}$ & $6.1\times 10^{39}$ \\
       SBGs \& GCs & $9.2\times10^{-6}$ & $3.5\times10^{40}$ &$3.5\times10^{41}$ & $6.1\times 10^{41}$ \\
       RQ AGN & $2.9\times 10^{-6}$ & $1.1\times 10^{41}$ & $1.1\times 10^{42}$ & $1.9\times 10^{42}$ \\
       GCs-acc & $1.0\times 10^{-6}$ & $3.2\times 10^{41}$ & $3.2\times 10^{42}$ & $5.6\times 10^{42}$\\
       RL AGN & $1.0\times 10^{-7}$ & $3.2\times 10^{42}$ & $3.2\times 10^{43}$ & $5.6\times 10^{43}$ \\
       BL Lacs & $4.9\times 10^{-9}$ & $6.6\times 10^{43}$ & $6.6\times 10^{44}$ & $1.2\times 10^{45}$ \\
       FSRQs & $2.1\times 10^{-12}$ & $1.5\times 10^{47}$ & $1.5\times 10^{48}$ & $2.7\times 10^{48}$\\
    \end{tabular}
    \caption{Steady-state source population parameters used for the benchmark source classes in the main text.  The effective local number density, $n_0$, for each class is obtained from \Refe~\cite{Murase:2016gly}.  The UHE neutrino luminosity, $L_\nu$, is fixed for each source class by demanding that the UHE neutrino flux from it saturates the diffuse UHE neutrino background.  Results are for our low, intermediate, and high benchmark diffuse UHE neutrino background; see Appendix~\ref{section:detailsonbackground}.}
    \label{tab:steady}
\end{table}

\begin{table}[t!]
    \centering
    \begin{tabular}{ccccc}
       Source class & $\mathcal{R}_0$~[Mpc$^{-3}$~yr$^{-1}$] & $E_\nu$~[erg]  & $E_\nu$~[erg]   & $E_\nu$~[erg] \\
        &  & (low $\nu$ bg.) & (interm. $\nu$ bg.) & (high $\nu$ bg.)\\
       \hline
       SNe \& newborn pulsars & $1.1\times10^{-5}$  & $9.0\times10^{47}$ & $9.0\times10^{48}$   & $1.6\times10^{49}$ \\
       Hypernovae & $2.3\times10^{-6}$ & $4.4\times10^{48}$ & $4.4\times10^{49}$   & $7.6\times10^{49}$  \\
       LL GRBs & $3.4\times10^{-7}$ & $3.0\times10^{49}$ & $3.0\times10^{50}$   & $5.3\times10^{50}$ \\
       HL GRBs & $1.0\times10^{-9}$ & $9.8\times10^{51}$ & $9.8\times10^{52}$   & $1.7\times10^{53}$  \\
       Jetted TDEs & $3.1\times10^{-10}$ & $3.3\times10^{52}$ & $3.3\times10^{53}$   & $5.7\times10^{53}$  \\
       \end{tabular}
    \caption{Transient source population parameters used for the benchmark source classes in the main text. The effective local burst rate, $\mathcal{R}_0$. for each class is from \Refe~\cite{Ackermann:2019ows} (see also \Refe~\cite{Murase:2018utn}).  The energy emitted as UHE neutrinos, $E_\nu$, is fixed for each source class by demanding that the UHE neutrino flux from it saturates the diffuse UHE neutrino background.  Results are for our low, intermediate, and high benchmark diffuse UHE neutrino background; see Appendix~\ref{section:detailsonbackground}.}
    \label{tab:transient}
\end{table}

For transient sources, the situation is reversed. If none is discovered, the brightest transients, with total energy per source larger than $10^{53}$~erg, would be disfavored: these include GRBs and TDEs, which are known candidates for UHECR acceleration.  Conversely, if even one transient source is discovered, it would be a strong indication in favor of these source classes. TDEs have a negative redshift evolution; however, we show in Appendix~\ref{section:effectofevolution} that accounting for their correct evolution does not  change results significantly. 

Our conclusions for steady-state sources are broadly unaffected by the size of the background diffuse neutrino flux: the precise parameter-space regions excluded depend on the background, but we find similar results for our low, intermediate, and high benchmarks. For transients, we find that a diffuse background much lower than $10^{-9}$~GeV~cm$^{-2}$~s$^{-1}$~sr$^{-1}$ would disfavor source discovery even for very bright sources. If the background diffuse neutrino flux is comparable to the projected 10-year IceCube-Gen2 sensitivity~\cite{IceCube-Gen2:2021rkf}, our conclusions are taken to their natural extreme.  Steady-state source discovery would disfavor all possible source classes; no discovery would disfavor none.  Transient source discovery would disfavor dim, abundant classes; no discovery would disfavor bright, rare ones.  On the other hand, if the diffuse flux is below the IceCube-Gen2 sensitivity, the possibility of observing point sources is evidently reduced dramatically. See Appendix~\ref{section:sourcepopulationvaryingbackground} for details. 


\section{Summary and outlook}  Discovering UHE neutrinos, beyond 100~PeV, is key to finding the origin of UHECRs.  We have shown that IceCube-Gen2 may discover them within 10 years of operation via searches for UHE multiplets.  Their discovery---and also their absence---would place powerful constraints on the population of UHE sources.  Our forecasts are state-of-the-art in the propagation and radio-detection of UHE neutrinos, and in the backgrounds that muddle source discovery. 

For steady-state sources, if even one is discovered, most candidate classes would be disfavored as individually dominant; if none is, only bright, rare sources would be disfavored.  For transient sources, if none is discovered, all candidate classes with total injected energy larger than $10^{53}$~erg would be disfavored as individually dominant.  In any case, comparable contributions from multiple source classes might remain viable~\cite{Bartos:2021tok}.

Our conclusions are broadly robust against different choices of the background level---currently unknown---the effective volume of the detector array---tentatively, contingent on further simulation work---and the detector energy resolution---since conservatively we do not use the event energy spectrum in our work---but are sensitive to the detector angular resolution.  We recommend upcoming UHE telescopes~\cite{Sasaki:2014mwa, ARA:2015wxq, PierreAuger:2016qzd, Adams:2017fjh, GRAND:2018iaj, Otte:2019knb, ARIANNA:2019scz, Aguilar:2019jay, Anchordoqui:2019omw, Deaconu:2019rdx, Nam:2020hng, Wissel:2020sec, IceCube-Gen2:2020qha, RNO-G:2020rmc, POEMMA:2020ykm, Horandel:2021prj} to target a zenith-angle resolution of about $2^\circ$.  Auspiciously, the requirements that grant IceCube-Gen2 sensitivity to source discovery via multiplets also grant it sensitivity to simultaneous measurement of the diffuse UHE neutrino flux and the neutrino-nucleon cross section~\cite{Valera:2022ylt, Esteban:2022uuw}. 

Finally, the identification of point sources belonging to specific classes will be greatly helped by comparison with specific source catalog and dedicated point source searches. Further, source identification could be eased by looking at the unbinned angular distribution of UHE neutrinos, rather than identifying multiplets in the binned angular distribution as we do here. 

In the next 10--20 years, we will have an opportunity to complete our picture of the high-energy Universe.  We provide advanced tools and forecasts to help realize it.


\smallskip

\acknowledgments
We thank Iv{\'a}n Esteban, Ke Fang, Christian Glaser, Kohta Murase, Foteini Oikonomou, and Walter Winter for useful discussion.  The authors are supported by the {\sc Villum Fonden} under project no.~29388.   This project has received funding from the European Union's Horizon 2020 research and innovation program under the Marie Sklodowska-Curie grant agreement No.~847523 ‘INTERACTIONS’.   This work used resources provided by the High Performance Computing Center at the University of Copenhagen.



\appendix


\section{Derivation of the probabilities to detect multiplets}
\label{section:appendixprobabilities}

\renewcommand{\theequation}{A\arabic{equation}}
\renewcommand{\thefigure}{A\arabic{figure}}
\setcounter{figure}{0}    

In the main text, we computed the size of the smallest multiplet required to claim detection of a source located at different positions in the sky, \ie, in each pixel in the sky.  For each pixel, we computed the global probability that one or more multiplets of that size appears anywhere in the sky as a result of random over-fluctuations of the diffuse background of UHE neutrinos and atmospheric muons.  From the global probability, we computed the local p-value, in each pixel, that is required to claim source detection.  Below, we derive the global probability and generalize it.
 
Reference~\cite{Fang:2016hop} adopted a different, but related strategy to ours.  They used the sky-wide distribution of the angular separation of detected neutrino pairs to reject the hypothesis of no point sources.  This strategy does not directly lead to locating point sources, but has the well-suited for identifying the presence of a large number of dim point sources that might not be discovered individually.  This is because a dim point source may produce a multiplet that is not large enough to claim lead to source discovery by itself, yet the detection of more than one such multiplet may reveal the subtle presence of point source, since it is unlikely that the pure diffuse background leads to sub-threshold fluctuations in the number of events in multiple pixels simultaneously.  Reference~\cite{Fang:2016hyv} applied these methods to locate sources by ranking the pairs according to their statistical significance.  Below, we extend our methods in a similar direction, using the combined information from more than one multiplet over all the sky.  However, we do not use this extension to produce the results in the main text.

In the $i$-th pixel, the probability of detecting a multiplet of any size at less than the local p-value $p$ is
\begin{equation}
 \pi_i(p)
 =
 \sum_{k=\bar{n}_i(p)}^{\infty}
 \frac{\mu_i^k}{k!} e^{-\mu_i} \;,
\end{equation}
where $\bar{n}_i(p)$ is the size of smallest multiplet required to claim detection at local p-value $p$, as defined in the main text.  If the number of events were a continuous variable, $\pi_i(p)$ would be identically equal to $p$. However, because the number of events is a discrete quantity, $\pi_i(p)$ jumps discontinuously and is always less than or equal to $p$.  

The probability $\pi_i(p)$ is local, in the sense that it involves a single pixel.  To shift to a global description, \ie, one that accounts for the full sky, we compute the probability for the detection of no multiplet anywhere in the sky, \ie,
\begin{equation}
 P_0(p)
 =
 \prod_{i=1}^{N_{\mathrm{pixels}}} \left(1-\pi_i(p)\right) \;,
\end{equation}
where $N_{\mathrm{pixels}}$ is the number of pixels that tessellate the sky.  The probability of detection of exactly one multiplet in the $j$-th pixel, and no multiplet in any other pixel, is
\begin{equation}
 P_{1;j}(p)
 =
 \pi_j(p)\prod_{i\neq j, i=1}^{N_{\mathrm{pixels}}}
 \left(1-\pi_i(p)\right) \;.
\end{equation}
From this, the probability of detection of exactly one multiplet in any of the pixels is
\begin{equation}
 P_1(p)
 =
 \sum_{j=1}^{N_{\mathrm{pixels}}}
 P_{1;j}(p) \;.
\end{equation}

These results can be generalized to the case of multiple multiplets: the probability of detection of $N_\mathrm{mult}$ multiplets in the pixels $j_1, \ldots, j_{N_\mathrm{mult}}$ is
\begin{equation}
 P_{N_\mathrm{mult};j_1,\ldots,j_{N_\mathrm{mult}}}(p)
 =
 \pi_{j_1} (p)\ldots\pi_{j_{N_\mathrm{mult}}}(p) 
 \prod_{i\neq (j_1,\ldots,j_{N_\mathrm{mult}}),i=1}^{N_{\mathrm{pixels}}}
 \left(1-\pi_i(p)\right) \;, 
\end{equation}
and the probability of detection of $N_\mathrm{mult}$ in any of the pixels is
\begin{equation}
 \label{equ:prob_geq_n}
 P_{N_\mathrm{mult}}(p)
 =
 \sum_{j_1=1}^{N_{\mathrm{pixels}}} 
 \sum_{j_2\neq j_1,j_2=1}^{N_{\mathrm{pixels}}}
 \ldots
 \sum_{j_{N_\mathrm{mult}}\neq (j_1,\ldots,j_{N_\mathrm{mult}-1}),j_{N_\mathrm{mult}}=1}^{N_{\mathrm{pixels}}}  P_{N_\mathrm{mult};j_1,\ldots,j_{N_\mathrm{mult}}}(p) \;.
\end{equation}
The functions $P_0(p)$, $P_1(p)$, \ldots, $P_{N_\mathrm{mult}}(p)$ represent how likely it is to detect $0$, $1$, \ldots, $N_\mathrm{mult}$ multiplets over all sky, each at local p-value $p$.  Using them, we define the probabilities $P_{\geq n}(p)$ of detecting at least $n$ multiplets in the pure-background case.  The probability $P_{\geq 0}(p) = 1$ identically.  The remaining probabilities are computed recursively via
\begin{equation}
 P_{\geq n}(p)
 =
 P_{\geq n-1}(p)-P_{n-1}(p) \;.
\end{equation}
The probabilities $P_{\geq n}(p)$ allow us to define the global p-values for the rejection of the pure-background hypothesis, accounting for the look-elsewhere effect.  In particular, if $n$ multiplets are detected, each with a local p-value smaller than $p$, the global p-value is $P_{\geq n}(p)$.  Indeed, the definition adopted for the global p-value in the main text coincides with $P_{\geq 1}(p)$.

\begin{figure}[t!]
    \centering
    \includegraphics[width=0.45\textwidth]{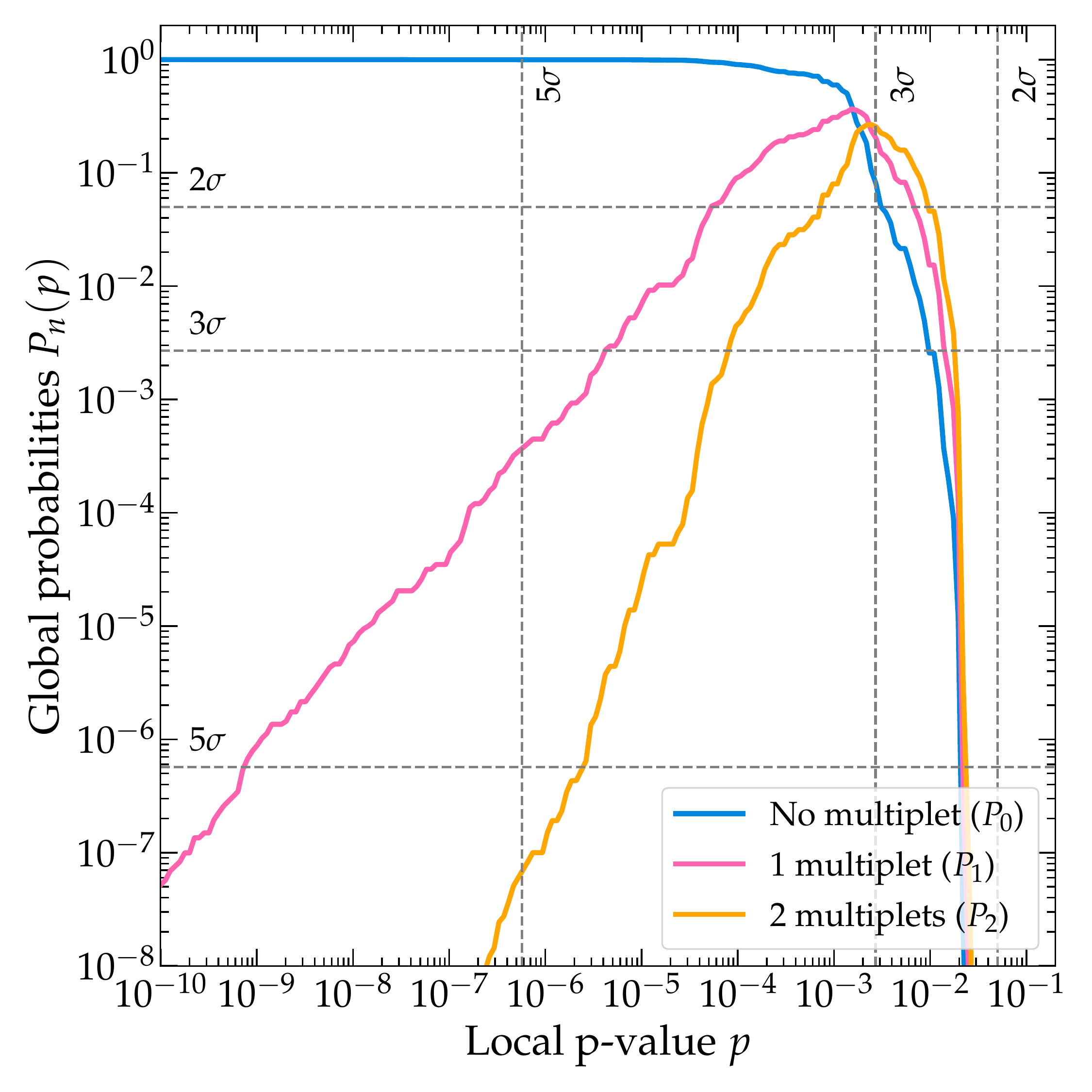}
    \includegraphics[width=0.45\textwidth]{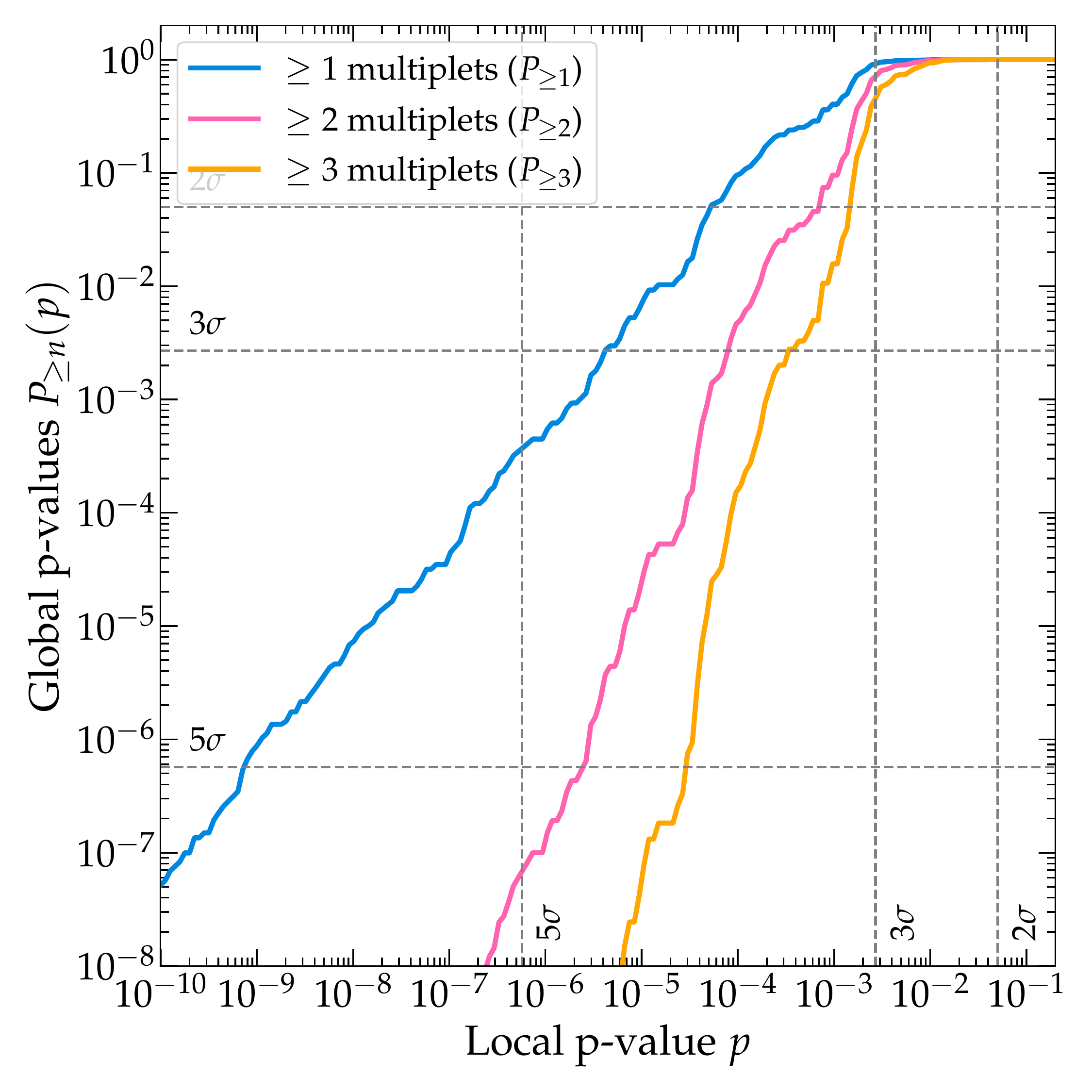}
    \caption{Global probabilities, $P_n(p)$ ({\it left}), and p-values, $P_{\geq n}$ ({\it right}), in the IceCube-Gen2 radio array, as functions of the local p-value $p$, for the first few multiplets.  The diffuse UHE neutrino background is our benchmark intermediate background; see Appendix~\ref{section:detailsonbackground}.  The radio array design is our baseline design (see main text), the detector angular resolution is our $\sigma_{\theta_z} = 2^\circ$, and the exposure time is $T = 10$~years.  See Appendix~\ref{section:appendixprobabilities} for details.}
    \label{fig:Pprobabilities}
\end{figure}

Figure~\ref{fig:Pprobabilities} shows the functions $P_n(p)$ and $P_{\geq n}(p)$ for the first few values of $n$, assuming our intermediate benchmark diffuse UHE neutrino background; see Appendix~\ref{section:detailsonbackground}.  For very small local p-values $p$, the probability of detecting a multiplet anywhere in the sky is negligible, \ie, $P_0(p) \approx 1$.  At larger values of $p$, it becomes increasingly likely that one or more multiplets are detected anywhere in the sky, due to the large number of pixels.  For our baseline choice of detector angular resolution of $\sigma_{\theta_z} = 2^\circ$ (see the main text), there are $N_{\mathrm{pixels}} \approx 3400$ pixels.  Hence, the threshold value of $p$ at which the probability of detecting  one multiplet, $P_1(p)$, becomes significantly large is between $10^{-4}$ and $10^{-3}$, of the order of $1/N_{\mathrm{pixels}}$.

Correspondingly, the global p-values for diffuse background rejection in the case of $1$, $2$, and $3$ detected multiplets, shown in \figu{Pprobabilities}, monotonically increase with $p$.  Further, the global p-value at a fixed value of $p$ decreases monotonically with the number of detected multiplets.  This means that, for a fixed global confidence level, if the sky contains more multiplets, then a larger value of $p$ is sufficient to claim background rejection. For example, claiming a $5\sigma$ detection with a single multiplet requires $p \approx 10^{-9}$, whereas claiming it with two multiplets requires only $p \approx 10^{-6}$.  Therefore, even if the two multiplets are detected with p-values that are insufficient to claim their discovery as two individual point sources, we can still claim that there is at least one point source in the sky.

Above, in \equ{prob_geq_n}, we implicitly assumed that, regardless of the number of multiplets detected, each multiplet was detected with the same p-value $p$.  However, in data collected by a real experiment, it is unlikely that all multiplets have the same local p-value.  In that situation, we can compute the global p-value for the hypothesis that no point source is present in any pixel as
\begin{equation}
 P_{\mathrm{global}}
 =
 \prod_{i=1}^{N_{\mathrm{pixels}}} 
 \sum_{k=n_i^{\mathrm{obs}}}^{\infty}
 \frac{\mu_i^k}{k!} e^{-\mu_i} \;,
\end{equation}
where $n_i^{\mathrm{obs}}$ is the observed number of events in the $i$-th pixel.  This approach amounts to a complementary, binned version of the method in \Refe~\cite{Fang:2016hop}.  However, it is only feasible a posteriori, with the data available.  That is why, in our analysis, we focus on a simplified version where all multiplets have the same local p-value.

\let\oldaddcontentsline\addcontentsline
\renewcommand{\addcontentsline}[3]{}
\subsection{Temporal look-elsewhere effect}
\let\addcontentsline\oldaddcontentsline

For transient sources, with a neutrino-emission period of duration $\delta t$, \eg, a flaring blazar or gamma-ray burst, this procedure requires modification. If the detector exposure time $T \gg \delta t$, a look-elsewhere effect may also happen in time, \ie, background over-fluctuations in any of the $N_{\delta t} \equiv T/\delta t$ time intervals may be misattributed to transient sources.  To prevent this, we raise the number of pixels to $N_{\rm pixels} \times N_{\delta t}$, scale the background event rate in each pixel by $1/N_{\delta t}$, and repeat the procedure introduced in the main text to claim source discovery.


\section{Radio-detecting ultra-high-energy neutrinos in IceCube-Gen2}
\label{section:appendixdetectionrate}

\renewcommand{\theequation}{B\arabic{equation}}
\renewcommand{\thefigure}{B\arabic{figure}}
\setcounter{figure}{0}

To compute neutrino-induced event rates at the IceCube-Gen2 radio array, we follow the procedure introduced in \Refe~\cite{Valera:2022ylt}.  Below, we sketch it; for details, we defer to \Refe~\cite{Valera:2022ylt}.

After UHE cosmic neutrinos reach the surface of the Earth, they travel underground towards the detector, IceCube-Gen2, at the South Pole, across different directions.  Underground, neutrinos interact with matter; as a result, their flux is attenuated and shifted to lower energies.  These effects grow with the matter column density traversed by the neutrinos on their way to the detector, \ie, they depend on the neutrino direction.  Downgoing neutrinos arrive at the detector from above, after traversing a small column density, and so are largely unattenuated.   Upgoing neutrinos arrive at the detector from below, after traversing up to diameter of the Earth, and so their flux is attenuated to the point of being negligible.  Earth-skimming neutrinos arrive at the detector from directions around the horizon, and so their flux is attenuated, but not fully; in general, it remains detectable.  

The leading neutrino-matter interaction at ultra-high energies is neutrino-nucleon ($\nu N$) deep-inelastic scattering (DIS)~\cite{CTEQ:1993hwr, Conrad:1997ne, Giunti:2007ry, Formaggio:2012cpf}.  An interaction can be neutral-current (NC)---when it is mediated by a $Z$ boson, \ie, $\nu_l + N \to \nu_l + X$, where $X$ are final-state hadrons--or charged-current (CC), \ie, $\nu_l + N \to l^- + X$---when it is mediated by a $W$ boson.  NC interactions shift neutrinos down to lower energies; CC interactions dampen the flux by removing neutrinos.  At ultra-high energies, CC interactions of $\nu_\tau$ may lead to ``$\nu_\tau$ regeneration," which makes $\nu_\tau$ more likely to survive their passage through Earth~\cite{Halzen:1998be, Becattini:2000fj, Beacom:2001xn, Dutta:2002zc, Yoshida:2003js, Bugaev:2003sw, Bigas:2008sw, Alvarez-Muniz:2018owm}.  We compute neutrino in-Earth propagation at next-to-leading order, for $\nu_\alpha$ and $\bar{\nu}_\alpha$ ($\alpha = e, \mu, \tau$) separately, using the state-of-the-art tool {\sc NuPropEarth}~\cite{Garcia:2020jwr, NuPropEarth}. For the internal matter density of Earth, we use the Preliminary Reference Earth Model~\cite{Dziewonski:1981xy}.

Inside the detector, a neutrino interacts with a proton or neutron of the Antarctic ice, via $\nu N$~DIS, triggering a high-energy particle shower that receives a fraction of the neutrino energy.  As the shower develops, the charged particles in it emit a coherent radio signal---Askaryan radiation~\cite{Askaryan:1961pfb}. It offers a promising method of detecting UHE neutrinos~\cite{Alvarez-Muniz:2001efi}, validated by laboratory measurements~\cite{Saltzberg:2000bk, Miocinovic:2006it, ANITA:2006nif} and  atmospheric shower measurements~\cite{PierreAuger:2014ldh, Schellart:2014oaa, Schroder:2016hrv, Huege:2016veh}.  Existing UHE neutrino telescopes---ARA~\cite{ARA:2015wxq}, ARIANNA~\cite{ARIANNA:2019scz}---and upcoming ones---RNO-G~\cite{Aguilar:2019jay}, IceCube-Gen2~\cite{IceCube-Gen2:2020qha}---target it.  Because radio waves have an attenuation length of roughly 1~km in ice~\cite{Barwick:2005zz}, compared to the roughly 100~m for optical light applicable to optical-Cherenkov detectors like IceCube, a large detection volume can be monitored with sparser instrumentation. The simulated detector volume is a cylinder 1.50~km tall, its top lid, with an area of 500~km$^2$, buried 100~m underground~\cite{Valera:2022ylt}. 

Following \Refe~\cite{Valera:2022ylt}, we model the IceCube-Gen2 response as the fraction of all the neutrino-initiated showers inside the geometric volume of the detector that trigger a signal and are recorded as events, and we use this fraction to define an effective volume for the experiment. This effective volume depends on the energy deposited in the ice and on the direction of the incoming neutrino. The simulations are performed in \Refe~\cite{Valera:2022ylt}, using the same tools as the IceCube-Gen2 Collaboration,  {\sc NuRadioMC}~\cite{Glaser:2019cws} and {\sc NuRadioReco}~\cite{Glaser:2019rxw}.  Appendix~\ref{section:experimentalconfiguration} details our baseline design of the radio array, and alternative designs.  Different array designs have different effective volumes; we compute the effective separately volume for each. The resulting effective volumes do not include the impact of reconstruction efficiencies, since this is not presently available in public literature.  This effect could somewhat reduce the total number of events and influence our results, but its importance is still under consideration.

Given a flux of $\nu_\alpha$ at the detector, $\Phi_{\nu_\alpha}^{\rm det}$, the differential rate of neutrino-induced events, in shower energy and direction, is
\begin{equation}
 \begin{split}
  \frac{d^2N_{\nu_\alpha}}{dE_{\rm sh} d\cos\theta_z}
  =& 
  2 \pi T n_t 
  \int_0^1 dy
  \left(
  \frac{E_{\nu_\alpha}^{\rm NC}(E_{\rm sh}, y)}{E_{\rm sh}}
  V_{{\rm eff}, \nu_\alpha}^{\rm NC}(E_{\rm sh}, \cos\theta_z) \right.
  \\
  &
  \left. \times
  \frac{d\sigma_{\nu_\alpha{\rm w}}^{\rm NC}(E_\nu, y)}{dy}
  \Phi^{\rm det}_{\nu_\alpha}(E_\nu,\cos\theta_z)
  \right\vert_{E_\nu = E_{\nu_\alpha}^{\rm NC}(E_{\rm sh}, y)}
  +~
  {\rm NC} \to \biggl. {\rm CC}
  \biggr)
   \;,
 \end{split}
 \label{equ:spectrum_true}
\end{equation}
where $T$ is the exposure time, $n_t \equiv N_{\rm Av}\rho_{\rm ice} / M_{\rm ice}$ is the number density of water molecules in ice, $N_{\rm Av}$ is Avogadro's number, $\rho_{\rm ice}$ is the density of ice, $M_{\rm ice}$ is the molar mass of water, and $\sigma_{\nu_\alpha{\rm w}}$ is the neutrino cross section on a water molecule.  The integration over $y$ accounts for the contribution of all possible initial neutrino energies $E_\nu$ that can produce a given shower energy $E_{\rm sh}$.  To account for the resolution of the detector in measuring shower energy and direction, we rewrite the differential event rate in terms of measured quantities, \ie, the reconstructed shower energy, $E_{\rm sh}^{\rm rec}$, and the reconstructed arrival direction, $\theta_z^{\rm rec}$.  We use energy and direction resolution functions, $\mathcal{R}_{E_{\rm sh}}$ and  $\mathcal{R}_{\theta_{z}}$, respectively, to cast the event rate in terms of measured quantities, \ie,
\begin{equation}
 \label{equ:spectrum_rec}
 \frac{d^2N_{\nu_\alpha}}
 {dE_{\rm sh}^{\rm rec} d\theta_{z}^{\rm rec}}
 =
 \int_{-1}^{+1} d\cos\theta_z
 \int_{0}^{\infty} dE_{\rm sh}
 \frac{d^2N_{\nu_\alpha}}{dE_{\rm sh} d\cos\theta_z} 
 \mathcal{R}_{E_{\rm sh}}(E_{\rm sh}^{\rm rec}, E_{{\rm sh}})~
 \mathcal{R}_{\theta_{z}}(\theta_{z}^{\rm rec}, \theta_z) \;.
\end{equation}
The energy and direction resolution functions are modeled as Gaussian functions centered at the real values of shower energy and direction, with widths $\sigma_{\theta_z}$ and $\sigma_{E_{\rm sh}}$, respectively.  To produce our main results, we set $\sigma_{\theta_z} = 2^\circ$ and $\sigma_{E_{\rm sh}} = 10^{\sigma_{\epsilon}} E_{\rm sh}$, where $\sigma_{\epsilon} = 0.1$ is the spread in the ratio $\epsilon \equiv \log_{10}(E^{\rm rec}_{\rm sh}/E_{\rm sh})$.  These values are chosen based on simulations~\cite{Anker:2019zcx, Glaser:2019rxw, Gaswint:2021smu, RNO-G:2021zfm, Stjarnholm:2021xpj, ARIANNA:2021pzm, Aguilar:2021uzt, Gaswint:2021smu, Stjarnholm:2021xpj, ARA:2021bss, Valera:2022ylt}.  To produce our results, we use the total event rate, summing \equ{spectrum_rec} over all flavors of neutrinos and anti-neutrinos.  To obtain the expected number of events in a given energy and direction bin, we integrate the total rate over the range reconstructed shower energy and direction boundaries of the bin.

\begin{figure}[t!]
    \centering
    \includegraphics[width=\textwidth]{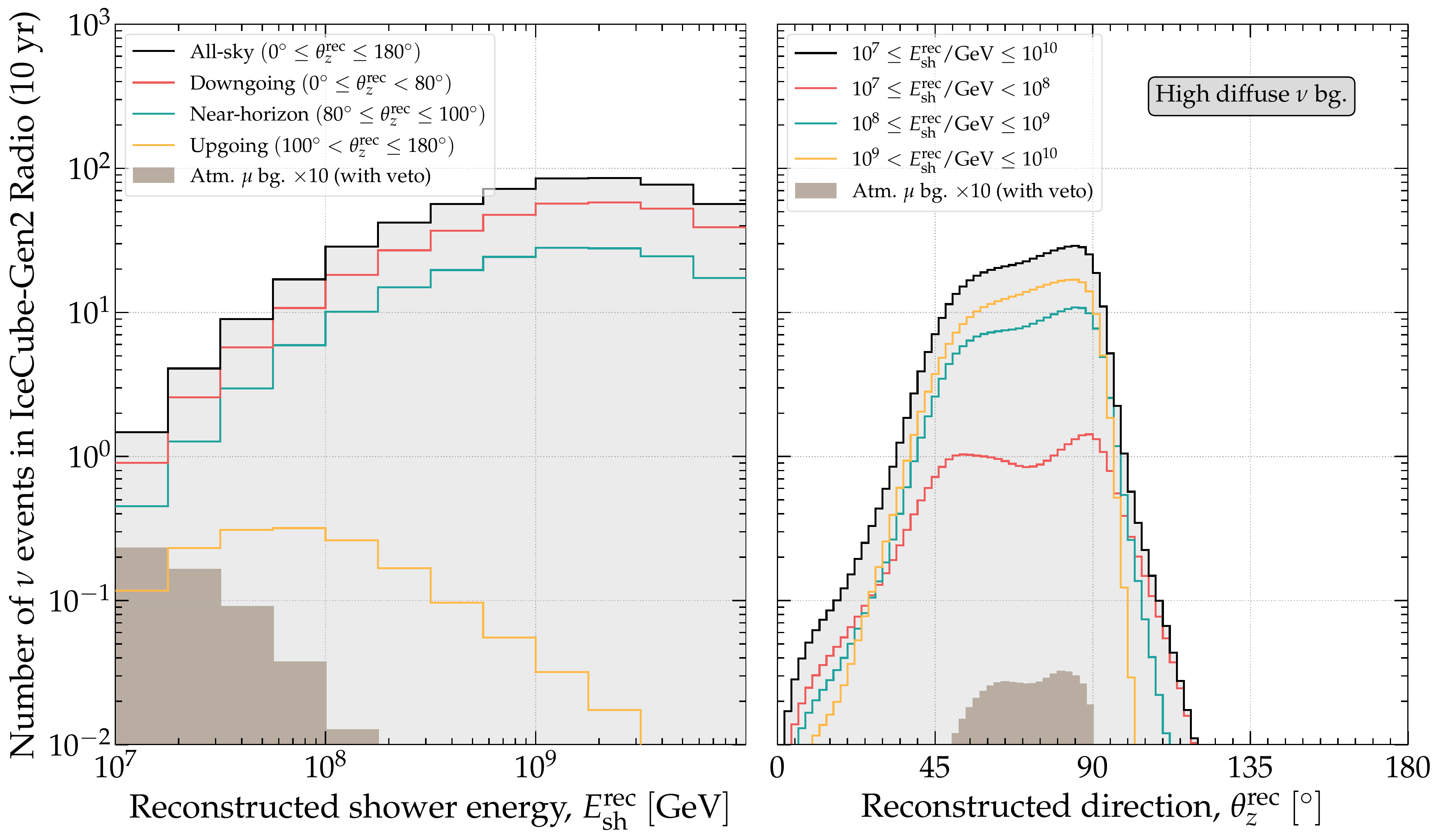}
    \caption{Distribution of events induced in IceCube-Gen2 by our high benchmark diffuse UHE neutrino diffuse model, in reconstructed shower energy ({\it left}) and direction ({\it right}), in $T = 10$~years of detector exposure time.  Events are computed following the procedure in Appendix~\ref{section:appendixdetectionrate}.  For this plot, we use our baseline design of the IceCube-Gen2 radio array, a detector shower energy resolution of $10\%$ in logarithmic scale, and a detector angular resolution in zenith angle of $\sigma_{\theta_z} = 2^\circ$.  See the main text and appendices \ref{section:appendixdetectionrate} and \ref{section:detailsonbackground} for details.}
    \label{fig:diffusebackgroundenergy}
\end{figure}

Figure~\ref{fig:diffusebackgroundenergy} illustrates the event rate, binned in reconstructed shower energy and direction, assuming for the neutrino flux our high benchmark flux; see the main text and Appendix~\ref{section:detailsonbackground}.  The event rates for our intermediate and low benchmarks are shifted-down versions of \figu{diffusebackgroundenergy}.  To produce our results, we do not use the distribution of events in energy, only in direction; see Appendix~\ref{section:detailsonbackground} for details.


\section{Impact of the detector angular resolution and the radio array design}
\label{section:experimentalconfiguration}

\renewcommand{\theequation}{C\arabic{equation}}
\renewcommand{\thefigure}{C\arabic{figure}}
\setcounter{figure}{0}    

In the main text, we presented results obtained using our choice of baseline detector angular resolution and of the design of the IceCube-Gen2 radio array.  The baseline detector angular resolution is $\sigma_{\theta_z}=2^\circ$ in zenith angle~\cite{Stjarnholm:2021xpj}.  The baseline design is composed of 144 hybrid stations (shallow + deep components) plus 169 shallow-only stations~\cite{IceCube-Gen2:2021rkf}; the deep components are buried 200~m underground.  In the main text, we commented on the effect on our results of using alternative choices of the angular resolution and the array design.  Here we present these results.

\begin{figure}[t!]
    \centering
    \includegraphics[width=0.47\textwidth]{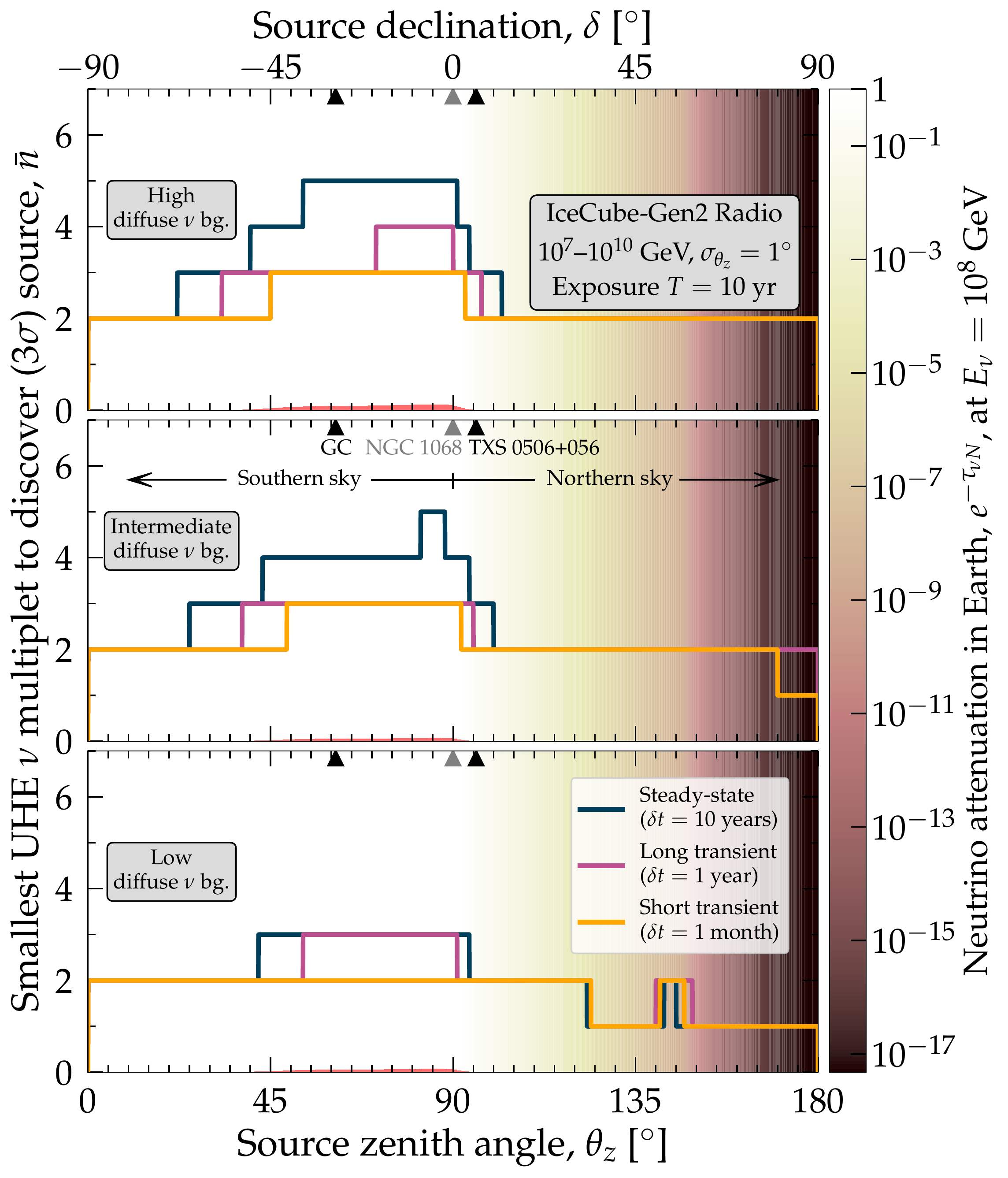}
    \includegraphics[width=0.47\textwidth]{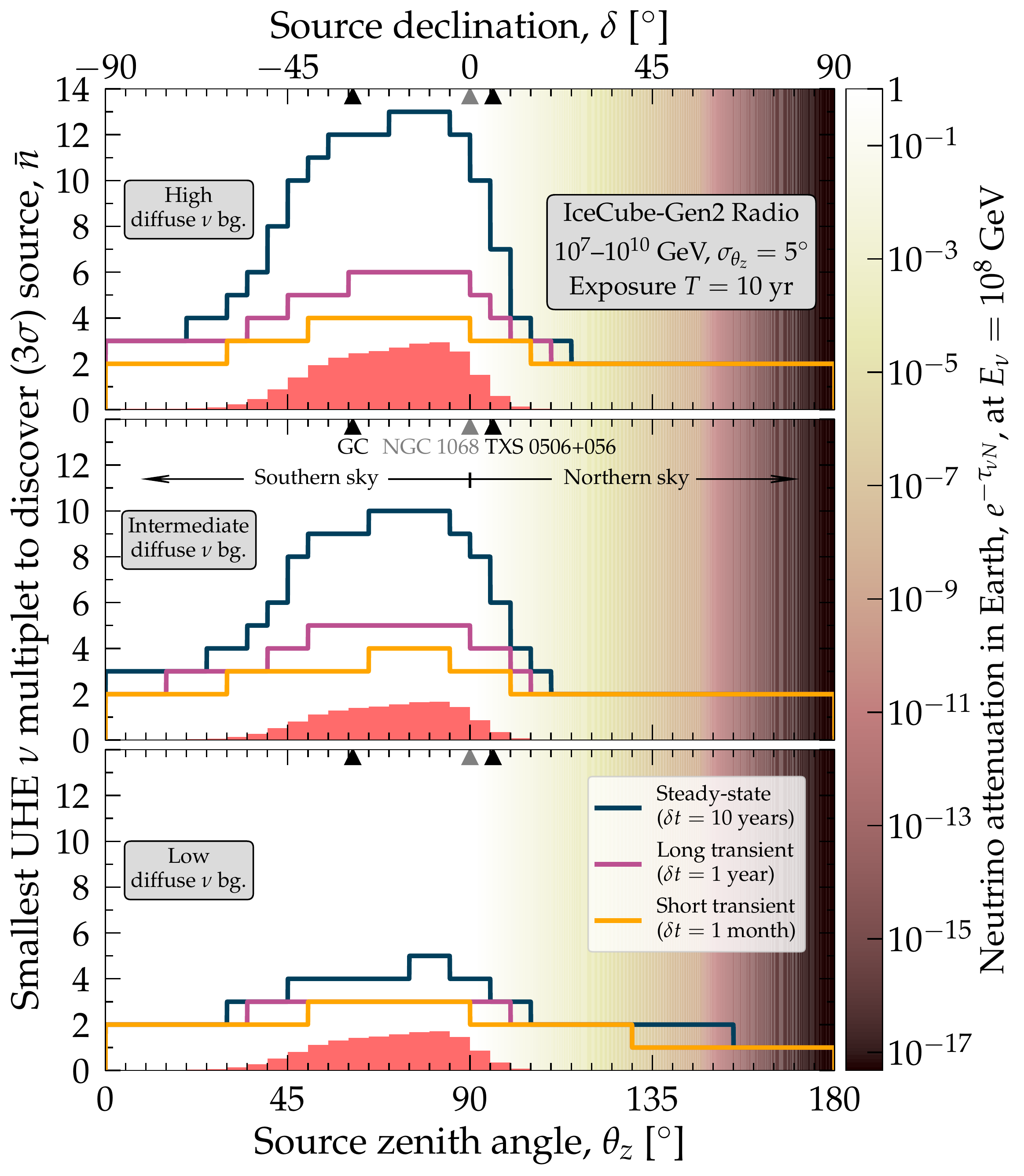}
    \includegraphics[width=0.47\textwidth]{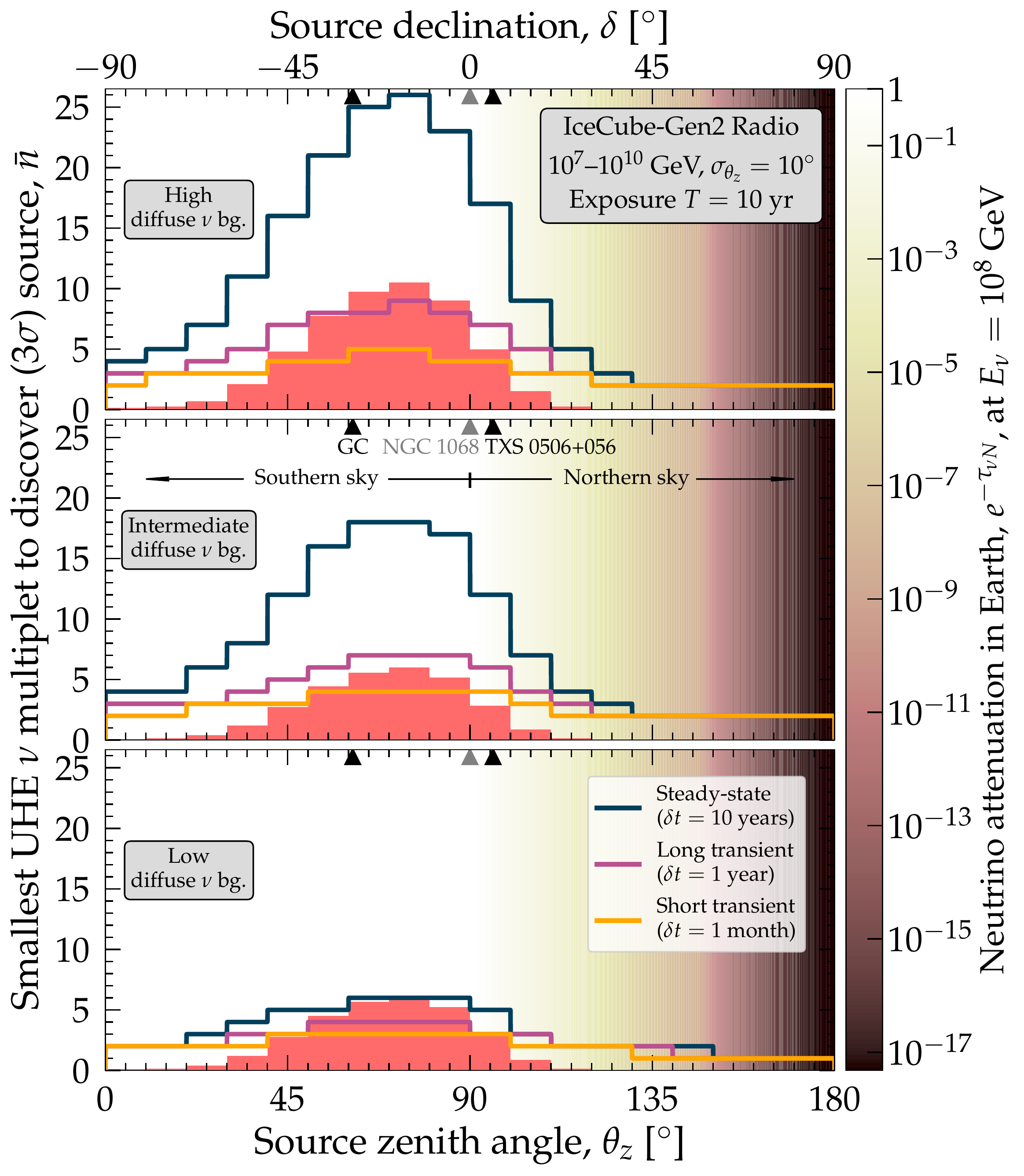}
    \caption{Same as Fig.~1 in the main text, but for different choices of the detector angular resolution: $\sigma_{\theta_z}=1^\circ$ ({\it top left}), $5^\circ$ ({\it top right}), and $10^\circ$ ({\it bottom}).}
    \label{fig:histogramdetectionfov}
\end{figure}

Figure~\ref{fig:histogramdetectionfov} shows the impact of varying the angular resolution on the smallest multiplet size needed to claim source discovery.  For comparison, in the main text we obtained results using $\sigma_{\theta_z} = 2^\circ$.  The angular resolution has a dramatic impact on the results: worsening it implies tessellating the sky with larger pixels and a corresponding larger diffuse background in each of them, hindering claims of source discovery.  For $\sigma_{\theta_z} = 5^\circ$, discovering steady-state sources would require up to size-13 multiplets if the diffuse background is high.  Transient sources would be more weakly affected: even for $\sigma_{\theta_z} = 10^\circ$, short-duration transients may require at  most a quintuplet or hexaplet to claim a detection.

\begin{figure}[t!]
 \centering
 \includegraphics[width=0.46\textwidth]{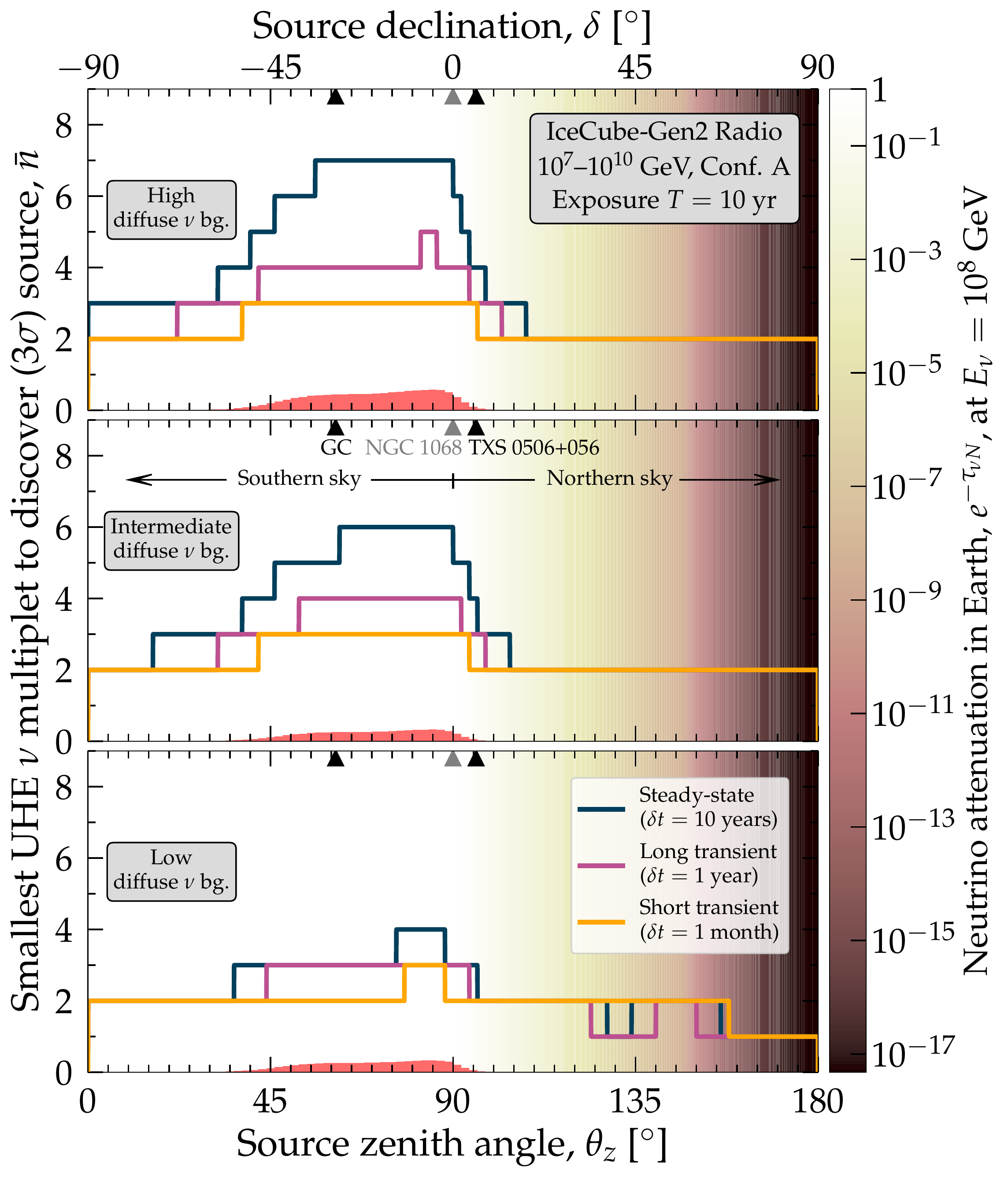}
 \includegraphics[width=0.46\textwidth]{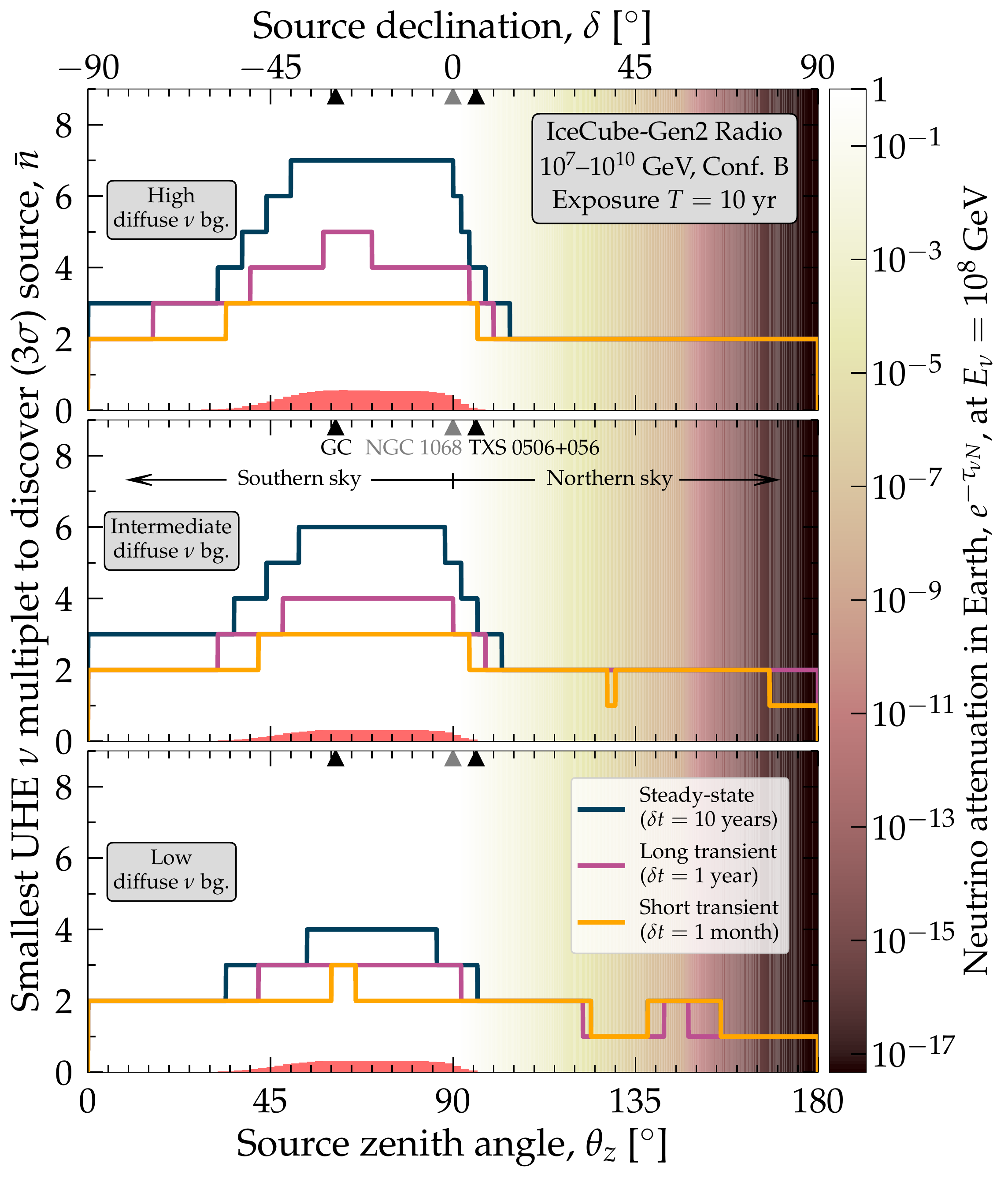}
 \caption{Same as Fig.~1 in the main text, but for two alternative designs of the IceCube-Gen2 radio array.  See Appendix~\ref{section:experimentalconfiguration} for details.}
 \label{fig:histogramdetectionconfigurations}
\end{figure}

Figure~\ref{fig:histogramdetectionconfigurations} shows the impact of using two alternative designs of the IceCube-Gen2 radio array on the  the smallest multiplet size to claim source discovery. 
We do not account for changes in the reconstruction efficiency of different array designs, since this information is presently not publicly available; see Appendix~\ref{section:appendixdetectionrate}.
In the main text we used our baseline array design.  Alternative array configuration A contains 361 shallow stations and 164 deep stations. Alternative array configuration B contains 208 hybrid stations only.  The prospects for detection are weakly dependent on the choice of array design; for our three design choices, at most heptaplets are needed to claim source discovery.  Because our exploration of alternative array designs is non-exhaustive, the weak dependence that we find should only be considered tentative.  More simulation work is necessary to claim this definitively.


\section{Diffuse ultra-high-energy neutrino  background}
\label{section:detailsonbackground}

\renewcommand{\theequation}{D\arabic{equation}}
\renewcommand{\thefigure}{D\arabic{figure}}
\setcounter{figure}{0}    

The diffuse UHE neutrino flux depends on the properties of UHECRs and their sources.  Because these are known uncertainly, the diffuse UHE neutrino flux is predicted uncertainly, and there are many competing theory models~\cite{Fang:2013vla, Padovani:2015mba, Fang:2017zjf, Heinze:2019jou, Muzio:2019leu, Rodrigues:2020pli, Anker:2020lre,  Muzio:2021zud}.  The diffuse UHE neutrino flux remains undiscovered, but there are upper limits on it, from IceCube~\cite{IceCube:2018fhm} and Auger~\cite{PierreAuger:2019ens}.

In our work, rather than adopting one of the above predictions for the diffuse UHE neutrino background, we adopt a flux that saturates, at each value of the neutrino energy, the IceCube upper limit~\cite{IceCube:2018fhm}.  This makes our background experiment-motivated rather than theory-motivated.  To account for future improvements in the detector sensitivity, we consider the three benchmark neutrino diffuse flux levels introduced in the main text---the present-day IceCube upper limit ({\it high}) and two smaller ones ({\it intermediate} and {\it low}).  See the main text for details. Because we do not use the energy distribution of the events, only their angular distribution, the specific shape of the neutrino energy spectrum that we adopt has no large consequence on our results.  What matters is only the total number of events, integrated across all energies.  The fact that our choice of spectral shape follows present-day upper limits and has no realistic theoretical counterpart does not impact our results significantly.

We take the diffuse neutrino flux to be isotropic at the surface of the Earth, in agreement with the expectation of a predominantly extragalactic origin. Neutrino propagation through the Earth makes the resulting distribution of neutrino-induced anisotropic; see Appendix~\ref{section:appendixdetectionrate} and \figu{diffusebackground}. For our three benchmarks, we assume approximate flavor equipartition at the surface of the Earth, \ie, equal proportion of $\nu_e+\bar{\nu}_e$, $\nu_\mu+\bar{\nu}_\mu$, and $\nu_\tau+\bar{\nu}_\tau$ in the flux.
This is the nominal expectations for high-energy neutrinos made in proton-proton and proton-photon interactions. Because in our analysis the neutrino-induced event rate is due to all flavors (see Appendix~\ref{section:appendixdetectionrate}) and we conservatively do not assume that flavor identification will be available, our results are only weakly sensitive to the precise flavor composition.  Further, we assume that the flux is equally divided between neutrinos and anti-neutrinos.  (At the highest energies, this may no longer be true, though we do not consider that possibility here; see, \eg, \Refs~\cite{Hummer:2010vx, Valera:2022ylt}).  At these energies, the neutrino and anti-neutrino cross sections are nearly equal, so uncertainties in the neutrino-to-antineutrino ratio have no discernible impact on the event rate.

\begin{figure}[t!]
    \centering
    \includegraphics[width=0.45\textwidth]{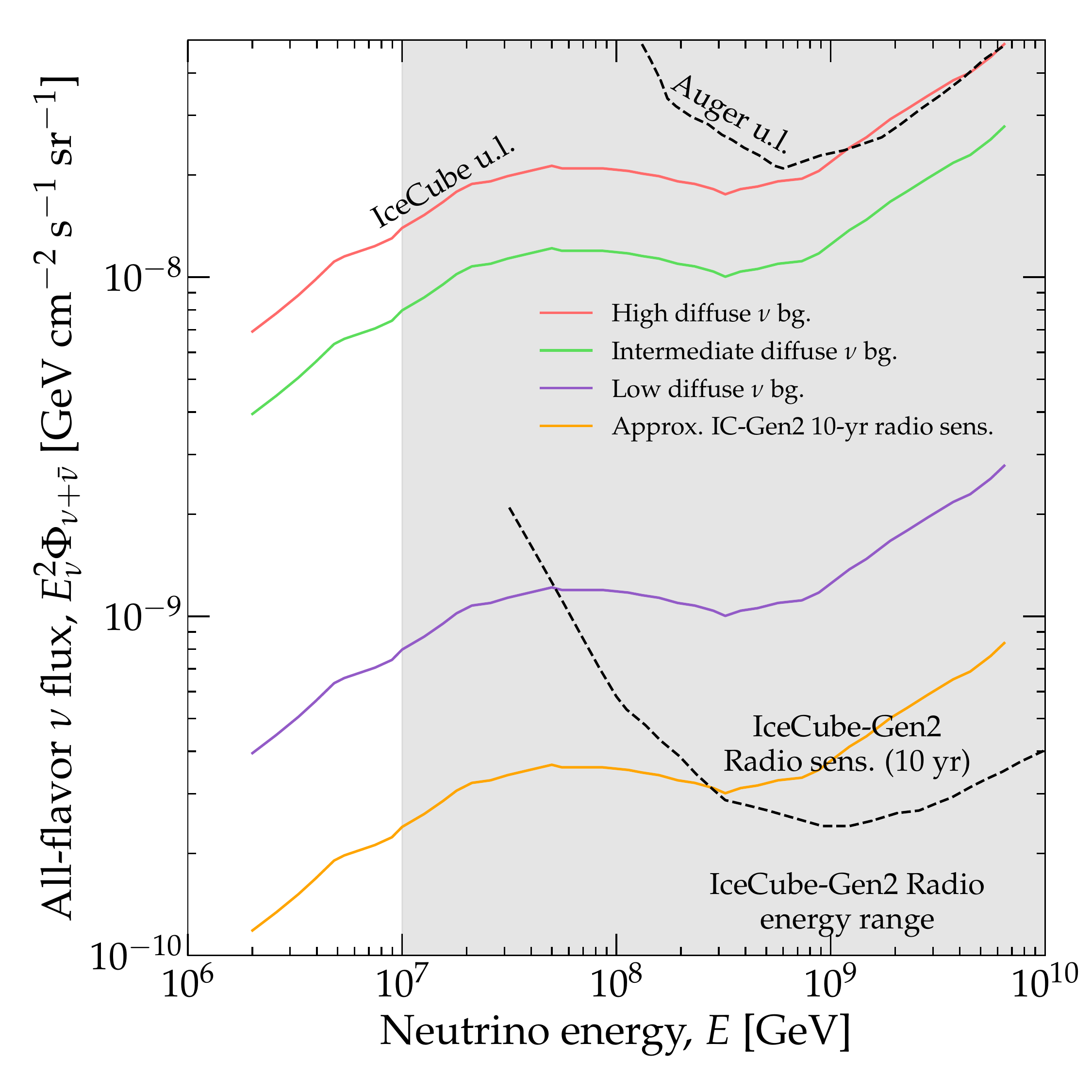}
    \includegraphics[width=0.45\textwidth]{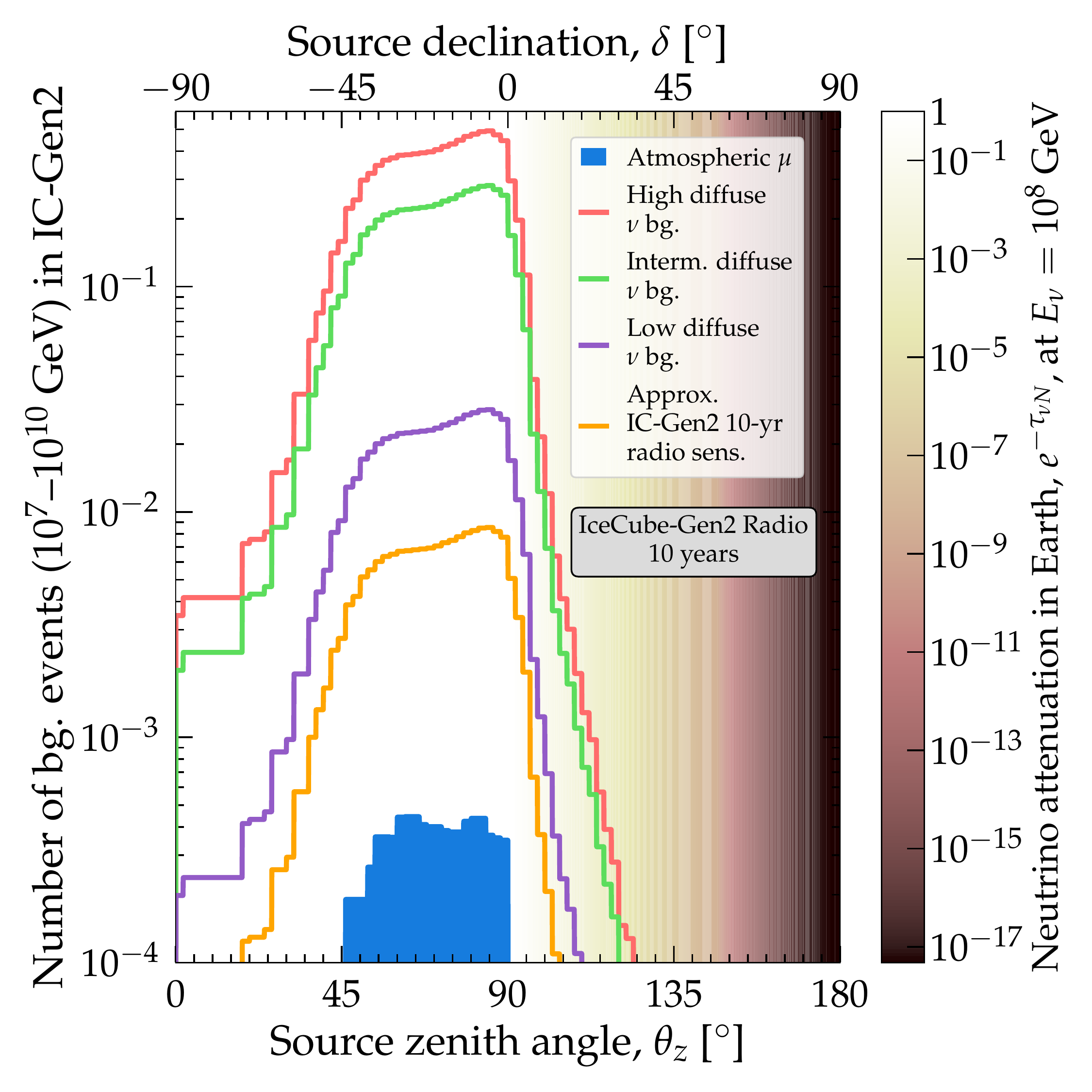}
    \caption{{\it Left:} Benchmark diffuse UHE neutrino flux models used in our work.  For comparison, we include the IceCube-Gen2 radio array sensitivity~\cite{IceCube-Gen2:2021rkf}, and the upper limit from Auger~\cite{PierreAuger:2019ens}.  
    {\it Right:} Angular distribution of the events induced by the benchmark diffuse neutrino background models in $T = 10$~years of IceCube-Gen2 exposure time.  For this plot, we use our baseline design of the IceCube-Gen2 radio array, and a detector angular resolution in zenith angle of $\sigma_{\theta_z} = 2^\circ$; see the main text.  The shading is the same as in Fig.~1 in the main text.  See Appendix~\ref{section:detailsonbackground} for details.}
    \label{fig:diffusebackground}
\end{figure}

Figure~\ref{fig:diffusebackground} shows our three benchmark diffuse UHE neutrino fluxes and their associated number of neutrino-induced events expected at IceCube-Gen2, in each pixel of the sky, computed following the procedure in Appendix~\ref{section:appendixdetectionrate}.  We include an additional benchmark flux that approximates the level of the projected 10-year sensitivity of the IceCube-Gen2 radio array~\cite{IceCube-Gen2:2021rkf}.  
Because diffuse flux is isotropic at the surface of the Earth, the angular distribution of the events reflects solely the effects of in-Earth propagation and the detector response.  The number of events close to the horizon is larger than at higher declination because there the detector angular response is strongest, while in-Earth attenuation is mild, compared to the severe attenuation experienced by upgoing neutrinos.

Figure~\ref{fig:diffusebackgroundenergy} shows details of the angular and energy distribution of events induced by our high benchmark diffuse UHE neutrino flux.  In our analysis, we do not use the energy distribution of the events; rather, we group all events in a single bin of reconstructed energy, $10^7$--$10^{10}$~GeV.  We show finer energy bins in \figu{diffusebackgroundenergy} for illustration only.  The energy dependence of the diffuse background model follows the energy spectrum of the IceCube upper limit, and is not theoretically motivated.  In terms of number of events, this leads to a signal that is more pronounced near $10^9$~GeV, mostly because of the higher effective volume of the detector at these energies. A different, possibly more realistic energy spectrum of the background diffuse neutrino flux would lead to a different event energy distribution. However, as pointed out earlier, because our analysis does not use the event energy distribution when looking for multiplets, our results are not significantly impacted by specific choices of the neutrino energy spectrum.

The second background that we account for is due to atmospheric muons coming from extensive air showers initiated by cosmic-ray interactions in the atmosphere~\cite{Garcia-Fernandez:2020dhb}. The number of events is highest at slant and near-horizontal directions, where the atmospheric muon flux is largest.  We use the same muon background as \Refe~\cite{Valera:2022ylt}, computed using the {\sc Sybill 2.3c} hadronic interaction model~\cite{Fedynitch:2018cbl}, mitigated by a surface veto, and subject to the same angular and energy resolution as neutrinos.  The background of muon-induced events is $< 0.1$ per year, sub-dominant to the background of neutrino-induced events. Even without the veto mitigating the atmospheric muon rate, we still expect less than one muon-induced event per year~\cite{Valera:2022ylt}.  Figures~\ref{fig:diffusebackground} and~\ref{fig:diffusebackgroundenergy} show the energy and angular distribution of events induced by atmospheric muons, subject to the same angular and energy resolution as neutrino-induced events.  For our benchmark background diffuse UHE neutrino fluxes, atmospheric muons are always sub-dominant in the total diffuse background, and influence weakly our results.  In principle, their impact could be further reduced by using the information on the energy spectrum in our analysis.  Since muons are produced by cosmic-ray interactions, their spectrum decreases steeply with energy, so their contribution is only relevant below $10^8$~GeV.  We conservatively do not use this information, but it could improve source-discovery prospects.

An additional background for UHE neutrino radio-detection may be due to the reflected cores of cosmic-ray showers~\cite{DeKockere:2022bto}. However, because its size is still under study, we cannot yet assess its impact on source discovery.


\section{Smallest multiplet size for $5\sigma$ detection and very short transients}
\label{section:results5sigma}

\renewcommand{\theequation}{E\arabic{equation}}
\renewcommand{\thefigure}{E\arabic{figure}}
\setcounter{figure}{0}    

\begin{figure}[t!]
    \centering
    \includegraphics[width=0.5\textwidth]{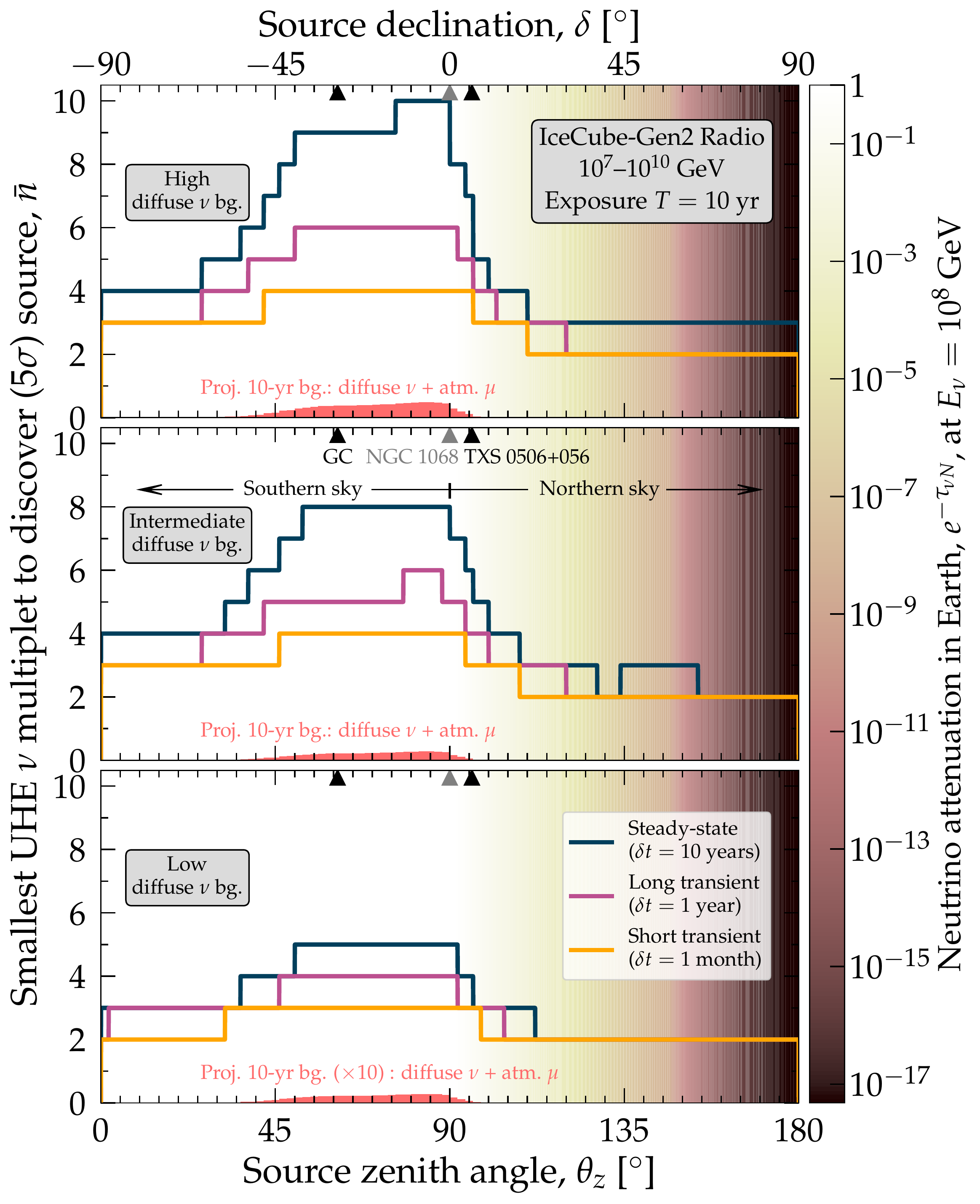}
    \caption{Same as Fig.~1 in the main text, but for source discovery at a global significance of $5\sigma$.}
    \label{fig:histogramdetection5sigma}
\end{figure}

Figure~1 in the main text shows the smallest multiplet size required to claim source discovery at a $3\sigma$ significance.  The largest multiplet, for steady-state sources, is a heptaplet.  

Figure~\ref{fig:histogramdetection5sigma} shows the smallest multiplet size required for a $5\sigma$ discovery.  In this case, larger multiplets are needed to overcome the background by a larger margin: up to decaplets may be needed to discover steady-state sources.

\begin{figure}[t!]
    \centering
    \includegraphics[width=0.46\textwidth]{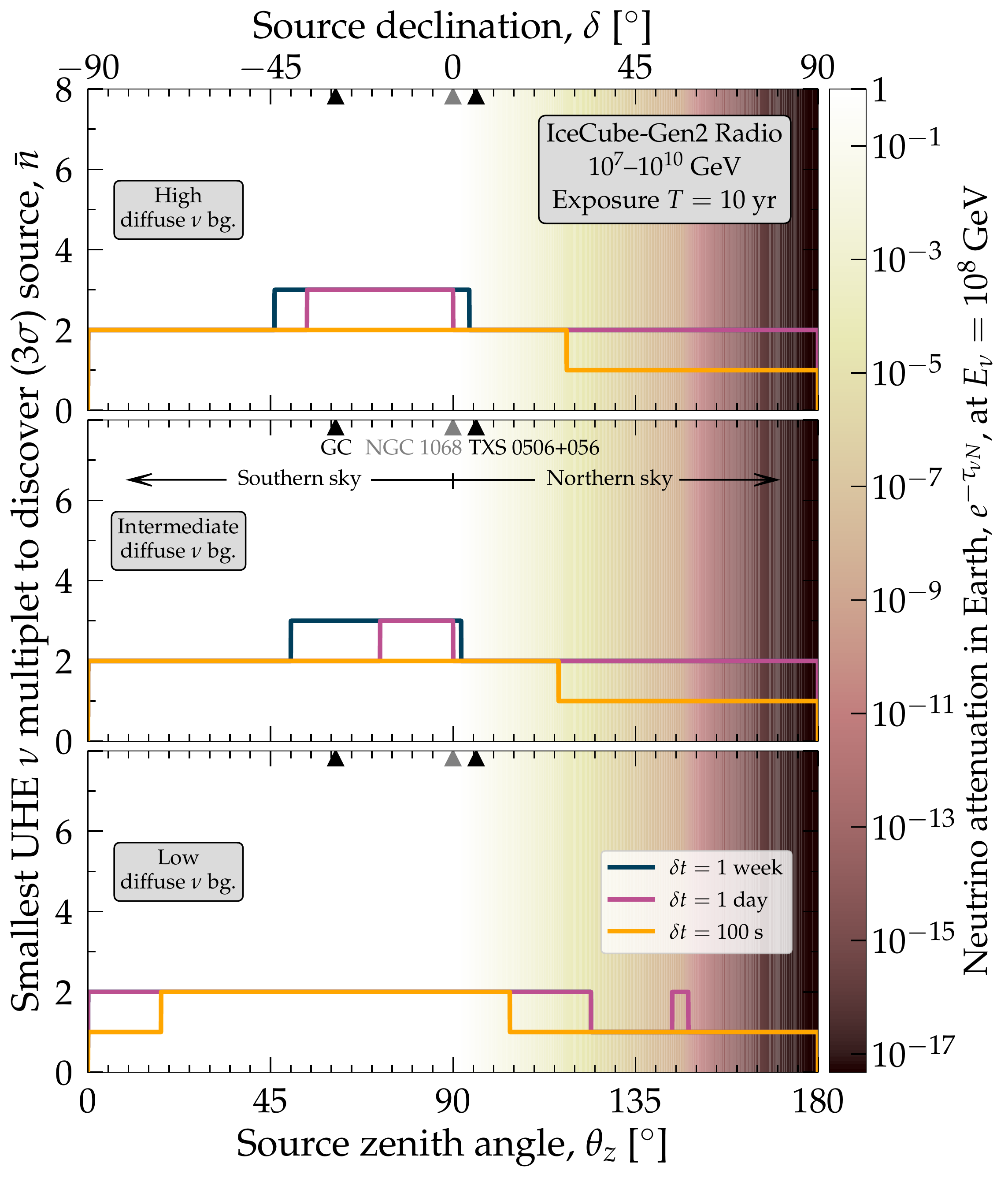}
    \includegraphics[width=0.46\textwidth]{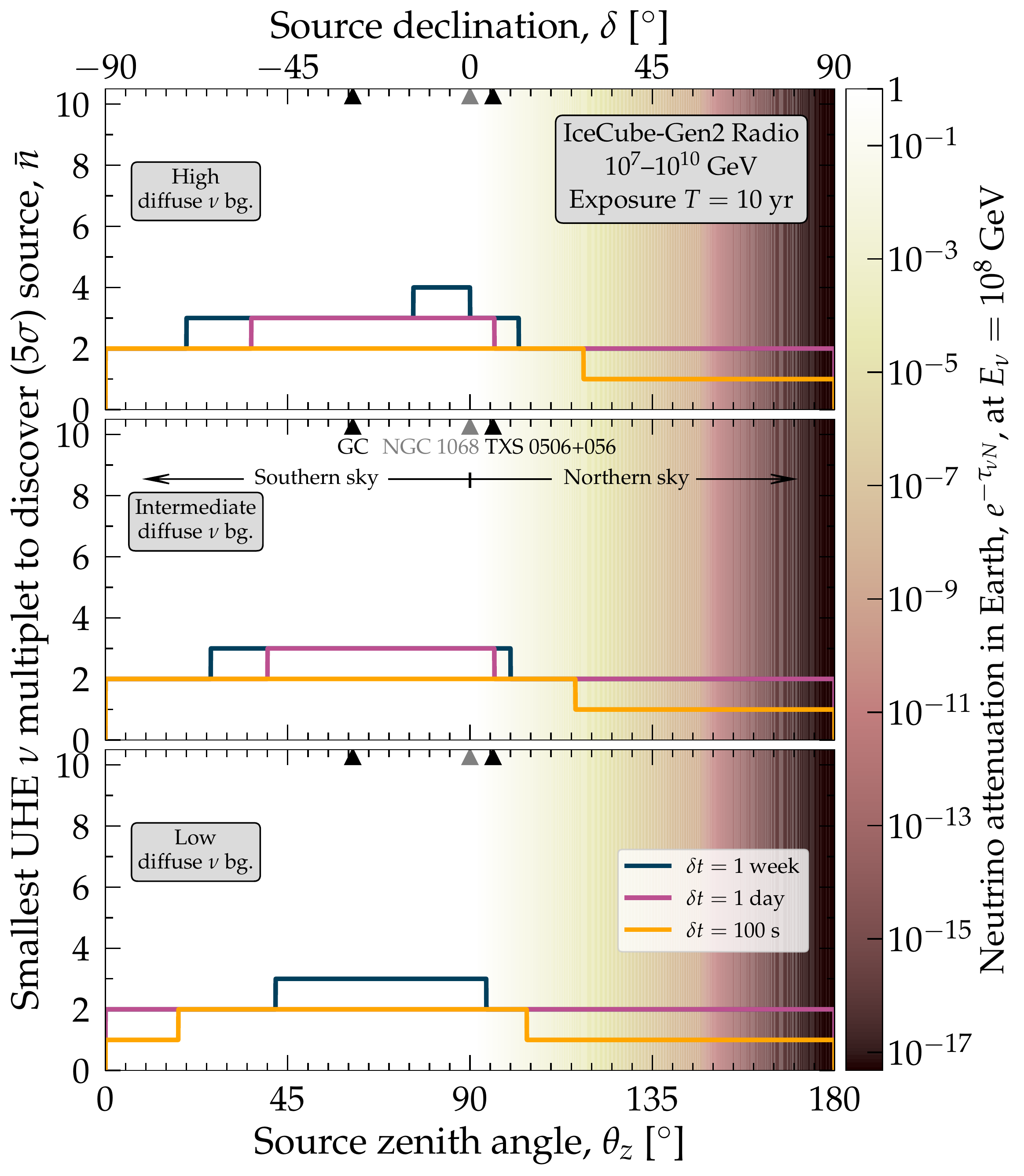}
    \caption{Same as Fig.~1 in the main text, but for very-short-duration transient sources, of duration $\delta t = 1$~week, 1~day, and 100~s. Results are for source discovery at $3\sigma$ ({\it left}) and $5\sigma$ ({\it right}) significance.  The diffuse background is too small to be visible and, therefore, is not shown.}
    \label{fig:histogramtransients}
\end{figure}

Figure~\ref{fig:histogramtransients} shows the smallest multiplet size needed to discover very-short-duration transient sources.  Because the background accumulated over the source duration is tiny, source discovery is significantly easier: doublets or triplets are sufficient anywhere in the sky.


\section{Constraints on source populations for varying background}
\label{section:sourcepopulationvaryingbackground}

\renewcommand{\theequation}{F\arabic{equation}}
\renewcommand{\thefigure}{F\arabic{figure}}
\setcounter{figure}{0}    

\begin{figure}[t!]
    \centering
    \includegraphics[width=0.45\textwidth]{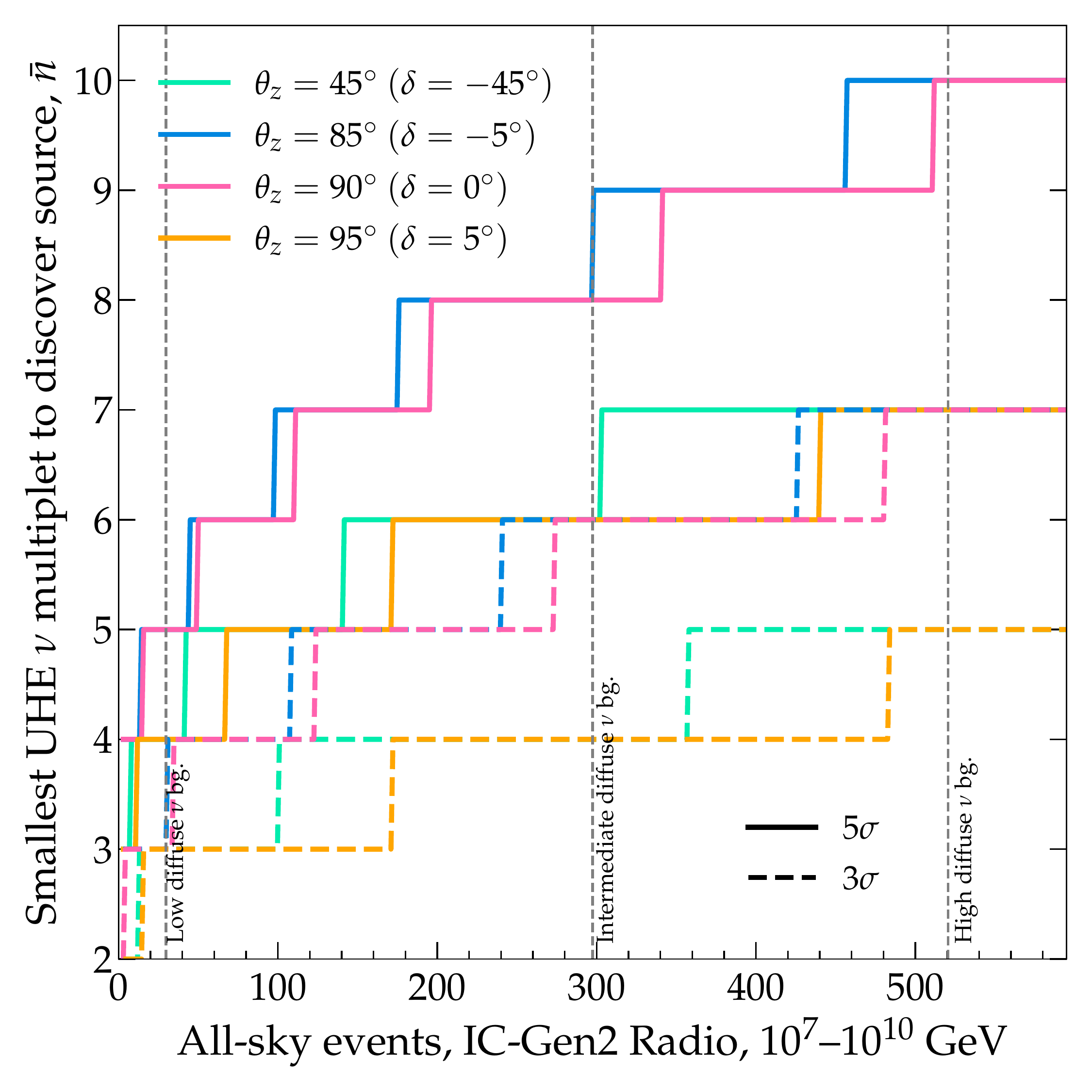}
    \caption{Threshold number of events for a detection as a function of the number of all-sky events. The predicted exposure time for the intermediate background model is shown in the top axis. The declination angle is fixed to four benchmark values of $\delta=-45^\circ$ (green), $\delta=-5^\circ$ (blue), $\delta=0^\circ$ (magenta), and $\delta=5^\circ$ (orange). We show the multiplet size for detection at $5\sigma$ (solid line) and $3\sigma$ (dashed line).}
    \label{fig:sizemultiplettime}
\end{figure}

Figure~\ref{fig:sizemultiplettime} shows how the size of the smallest multiplet needed to claim source discovery varies with the all-sky number of background-induced events, which is proportional to the background diffuse UHE neutrino flux.  A lower background means that smaller multiplets are needed to claim source discovery.

\begin{figure}[t!]
    \centering
    \includegraphics[width=0.45\textwidth]{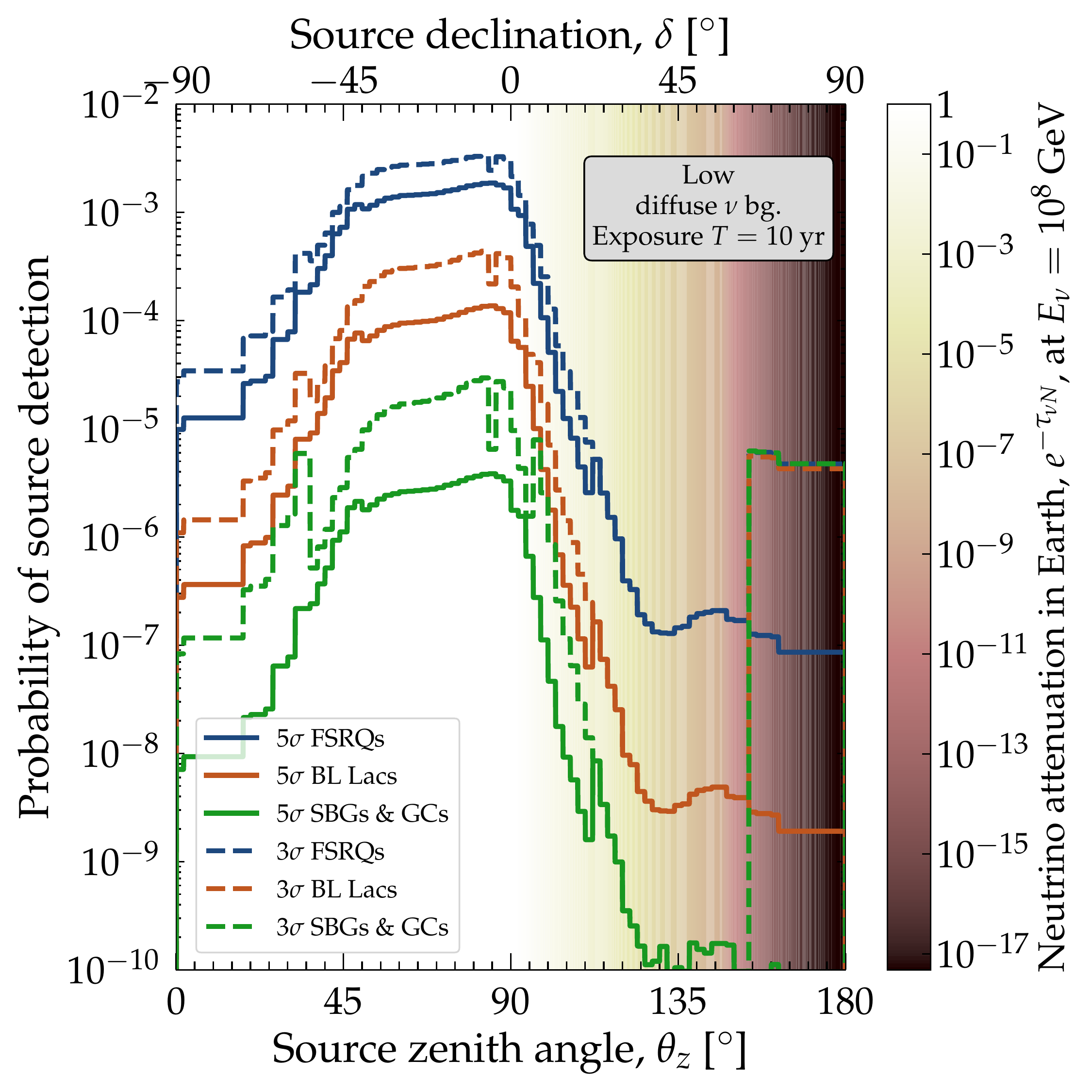}
    \includegraphics[width=0.45\textwidth]{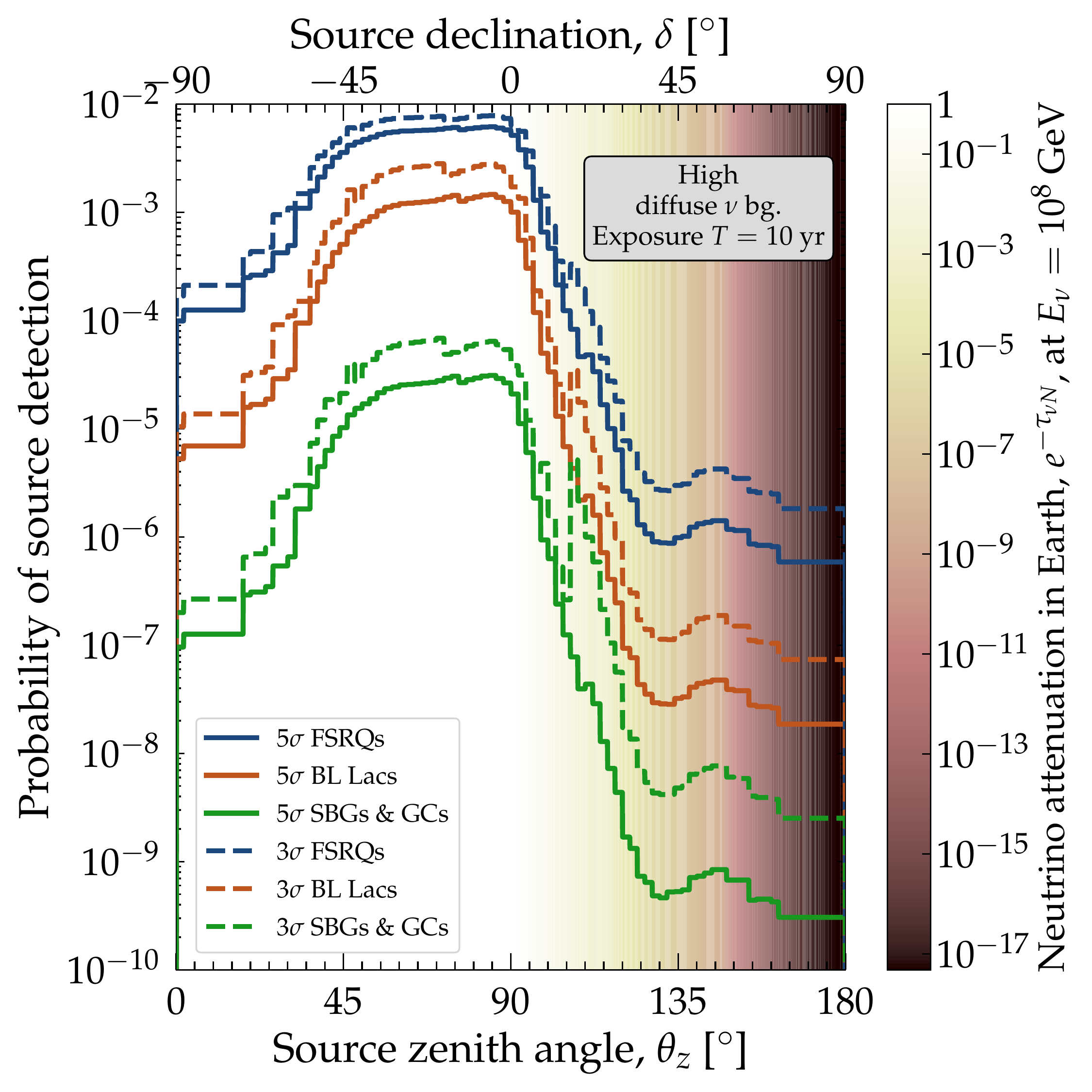}
    \caption{Same as \figu{detectionprobability}, but using our benchmark low ({\it left}) and high ({\it right}) UHE neutrino background.  See Appendix~\ref{section:sourcepopulationvaryingbackground} and the main text for details.}
    \label{fig:detectionprobabilityvaryingbackground}
\end{figure}

Figure~\ref{fig:detectionprobabilityvaryingbackground} shows the probability of source discovery in each pixel, for our low and high benchmarks of the diffuse UHE neutrino background.  \figu{detectionprobability} shows the case for our intermediate benchmark.  A larger diffuse background requires larger multiplets to claim source discovery.  However, because in Figs.~\ref{fig:detectionprobability} and \ref{fig:detectionprobabilityvaryingbackground} the neutrino flux from each source class is chosen to saturate the diffuse background (see Tables~\ref{tab:steady} and \ref{tab:transient}), a larger diffuse background implies a higher chance of source discovery in every pixel.

\begin{figure}[t!]
 \centering
 \includegraphics[width=0.45\textwidth]{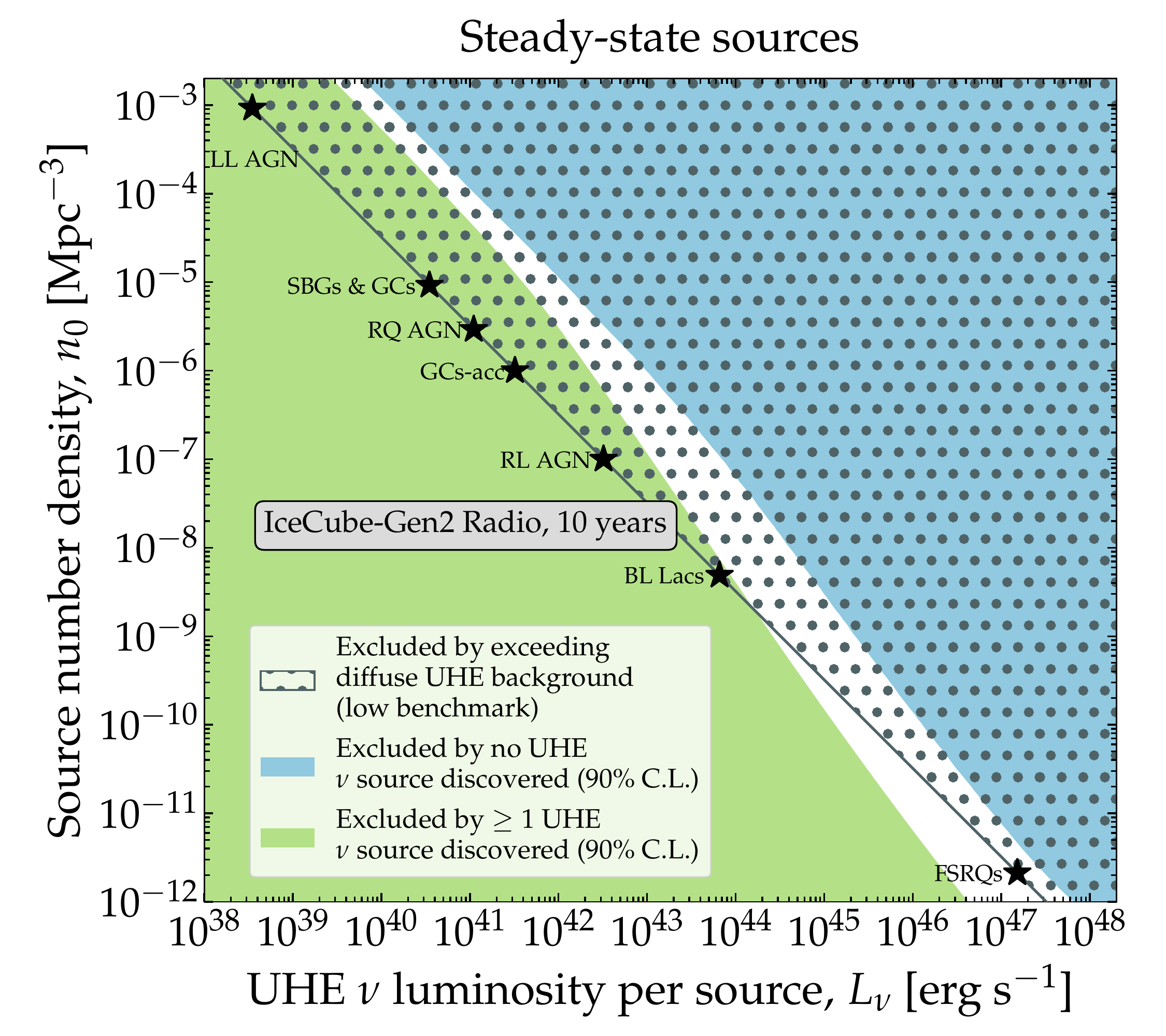}
 \includegraphics[width=0.46\textwidth]{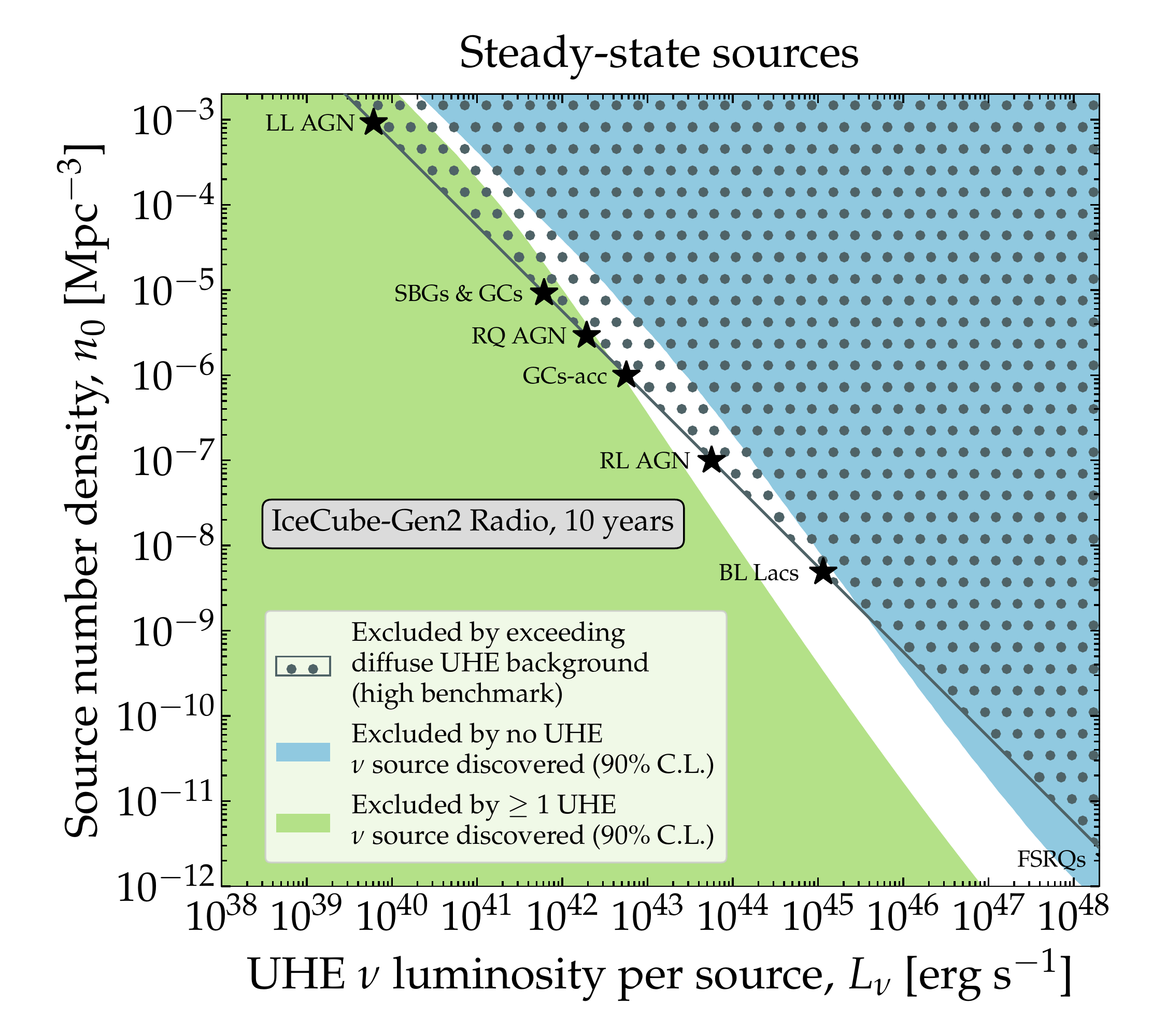}
 \includegraphics[width=0.45\textwidth]{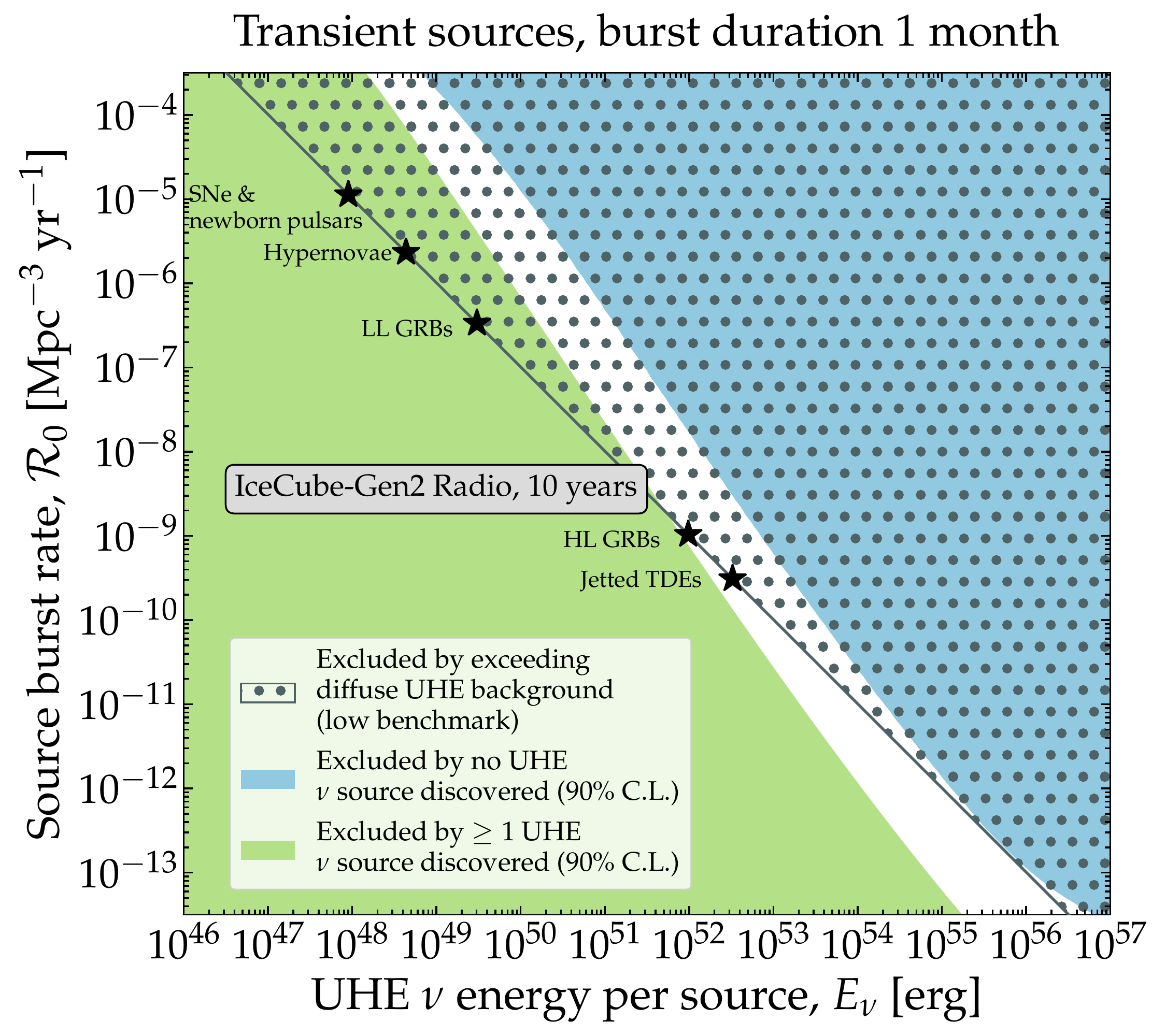}
 \includegraphics[width=0.46\textwidth]{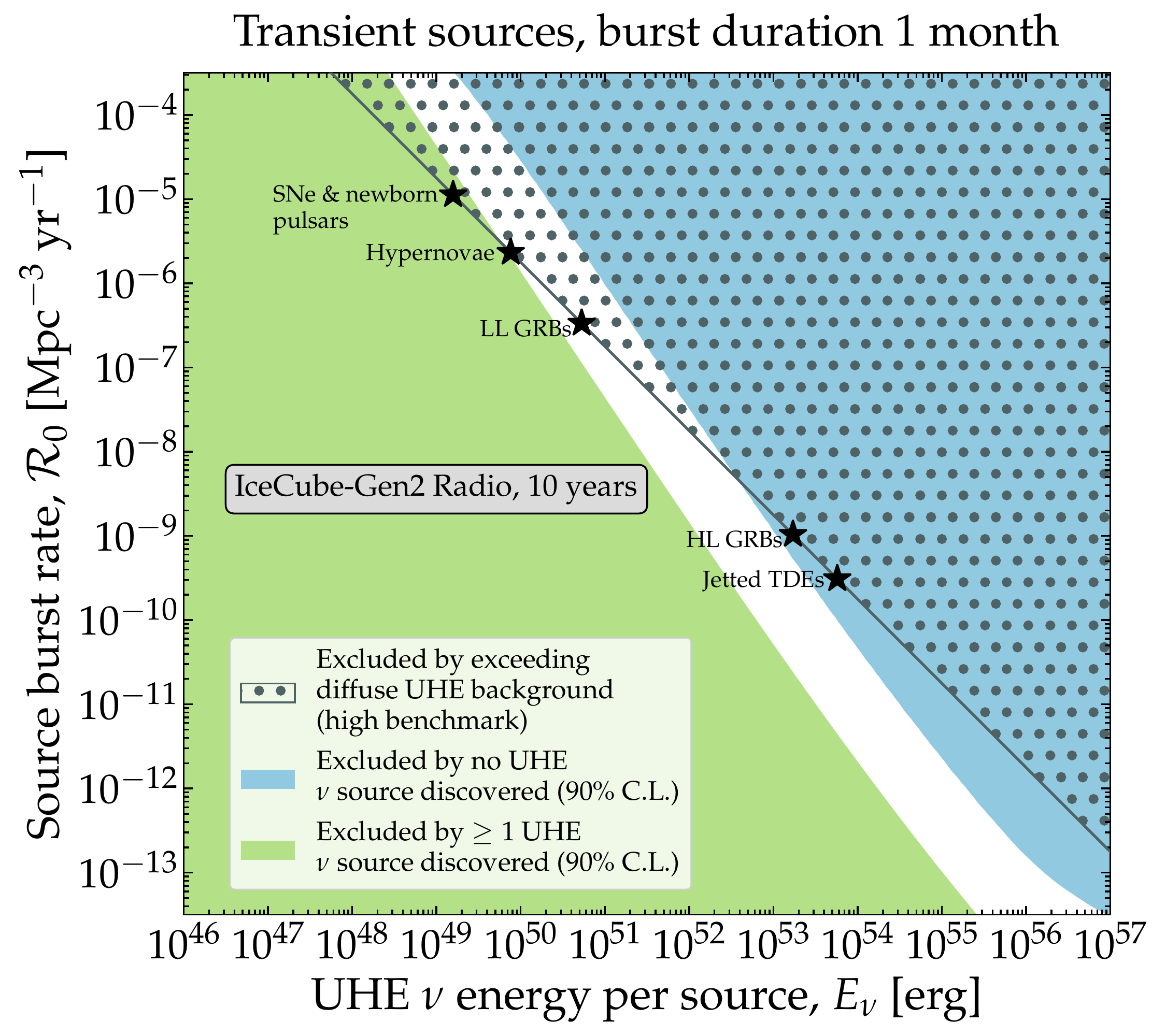}
 \caption{Same as Fig.~2 in the main text, but using our benchmark low ({\it left column}) and high ({\it right column}) UHE neutrino background, for steady-state sources ({\it top row}) and transient sources ({\it bottom row}).  See the main text and Appendix~\ref{section:sourcepopulationvaryingbackground} for details.}
 \label{fig:pointsourcesensitivityvaryingbackground}
\end{figure}

Figure~\ref{fig:pointsourcesensitivityvaryingbackground} shows the constraints on UHE neutrino sources, from the discovery or absence of UHE multiplets, for our benchmark low and high UHE neutrino background.  In the main text, Fig.~2 shows constraints for our intermediate benchmark. Fig.~\ref{fig:pointsourcesensitivityvaryingbackground} shows that the qualitative impact of the choice of background on the constraints is mild.  For steady-state sources, regardless of the background, if even a single source is discovered, most known candidate source classes would be disfavored.  Figure~\ref{fig:pointsourcesensitivityvaryingbackground} reveals that this is especially true if the background is low, because then sources cannot be too bright without exceeding it.  For transient sources, in the high-background case our conclusions are unchanged, whereas in the low-background case the absence of a detection would not disfavor the known source classes.

\begin{figure}[t!]
 \centering
 \includegraphics[width=0.502\textwidth]{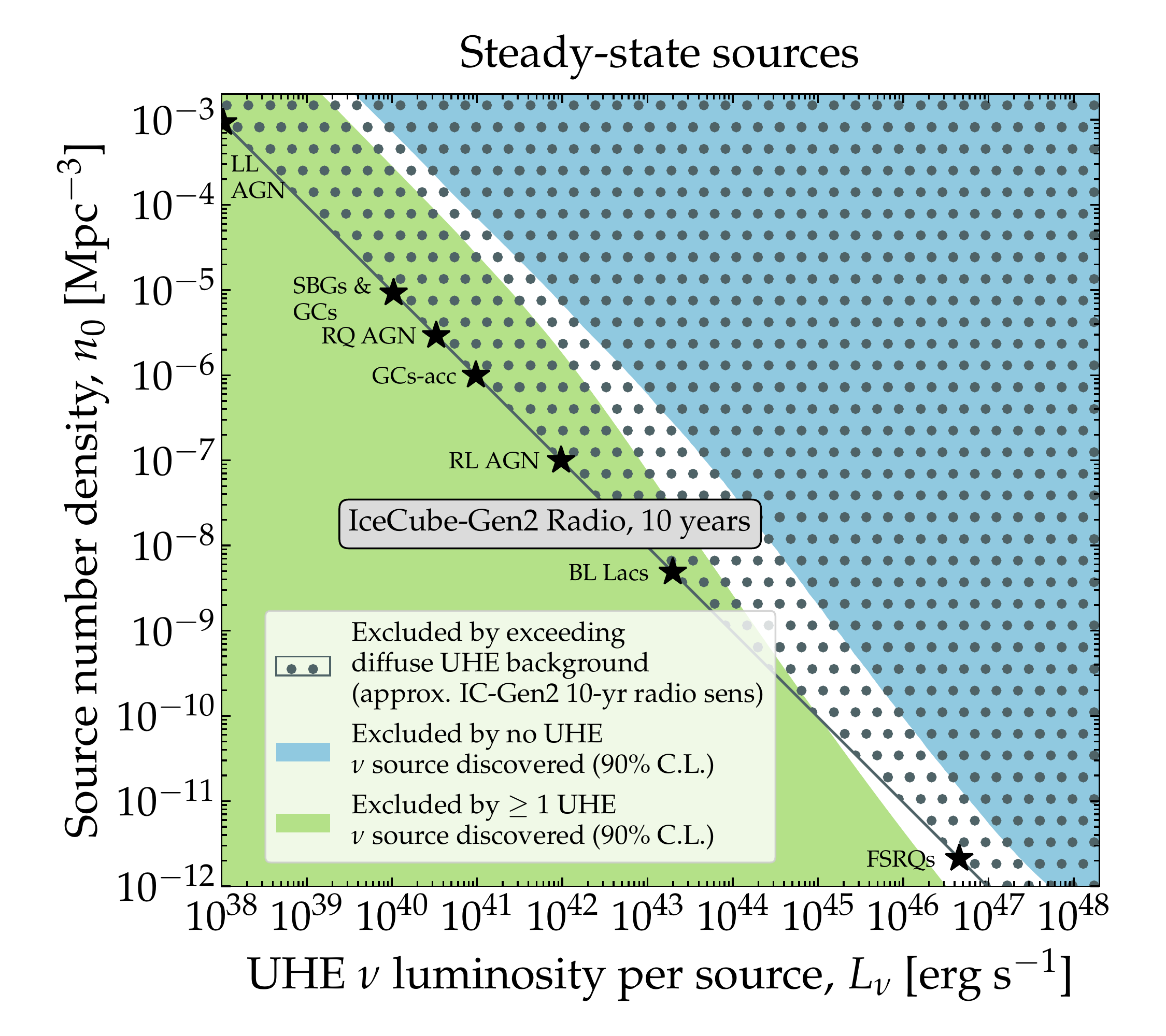}
 \includegraphics[width=0.485\textwidth]{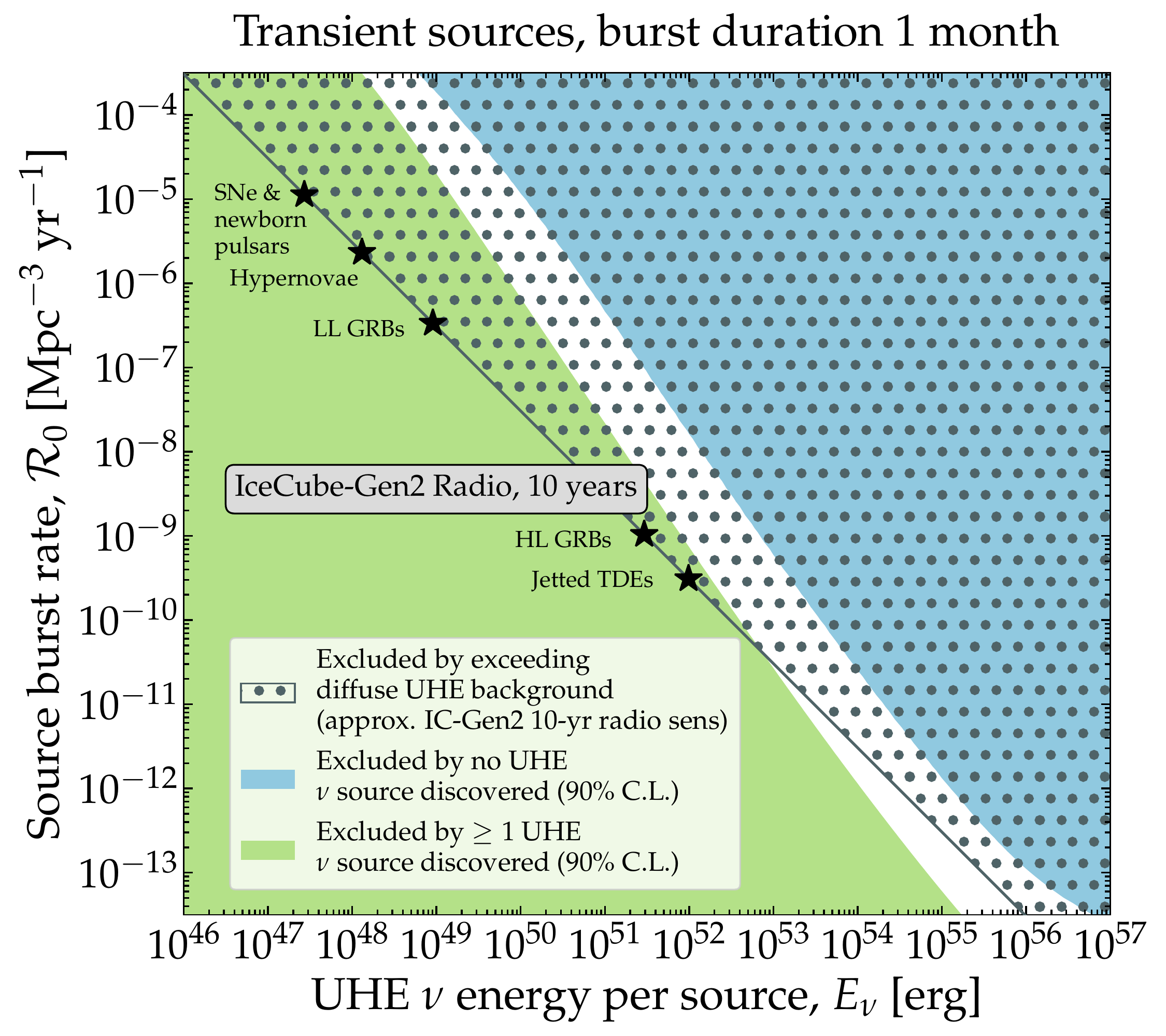}
 \caption{Same as Fig.~2 in the main text, but using an UHE neutrino background approximately coinciding with the IceCube-Gen2 radio array sensitivity~\cite{IceCube-Gen2:2021rkf}.  See Appendix~\ref{section:sourcepopulationvaryingbackground} and the main text for details.}
 \label{fig:pointsourcesensitivitylowestbg}
\end{figure}

Figure~\ref{fig:pointsourcesensitivitylowestbg} shows the constraints on the source population assuming an even lower background, one approximately at the same level as the projected 10-year sensitivity of the IceCube-Gen2 radio array sensitivity~\cite{IceCube-Gen2:2021rkf}; see \figu{diffusebackground}. For such a low background, we expect about 9 events all-sky in 10 years of IceCube-Gen2 for our baseline array design.  Indeed, in the case of no point-source discovery, \figu{pointsourcesensitivitylowestbg} shows that point source discovery is not expected for any source class. It is entirely possible that the diffuse neutrino flux is lower even than our low benchmark model.  In that, case the bounds obtained from the absence of point sources would be significantly weaker than the requirement of not exceeding the bounds on the diffuse neutrino flux. 


\section{Impact of redshift and luminosity evolution}
\label{section:effectofevolution}

\renewcommand{\theequation}{G\arabic{equation}}
\renewcommand{\thefigure}{G\arabic{figure}}
\setcounter{figure}{0}  

\begin{figure*}[b!]
 \centering
 \includegraphics[scale=0.5]{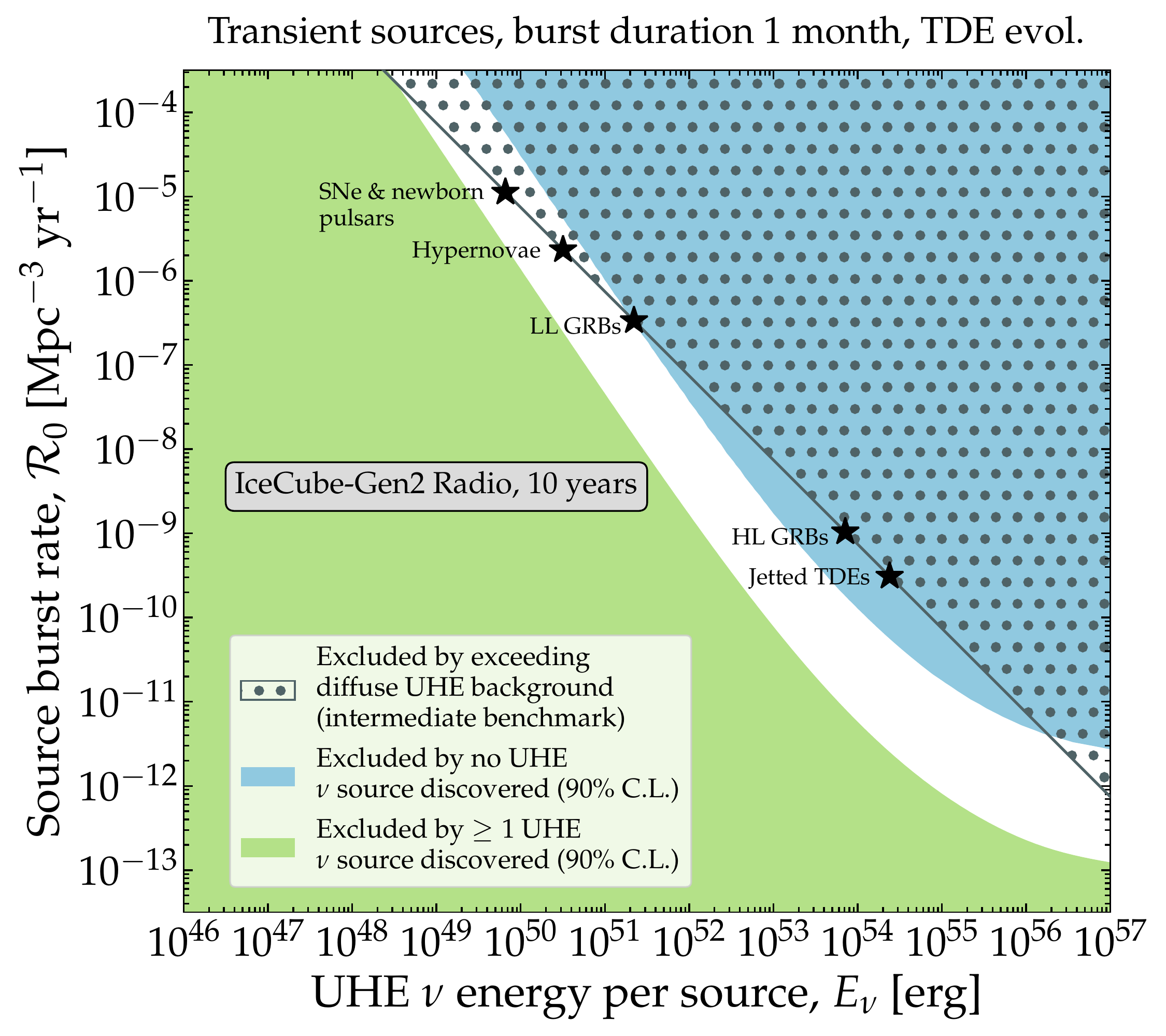}
 \caption{Same as \figu{pointsourcesensitivity} in the main text, for transient sources, but assuming a TDE-like redshift evolution for the number source density.  See Appendix~\ref{section:effectofevolution} for details.  }
 \label{fig:pointsourcesensitivityjettedtde}
\end{figure*}

\begin{figure*}[t!]
 \centering
 \includegraphics[scale=0.5]{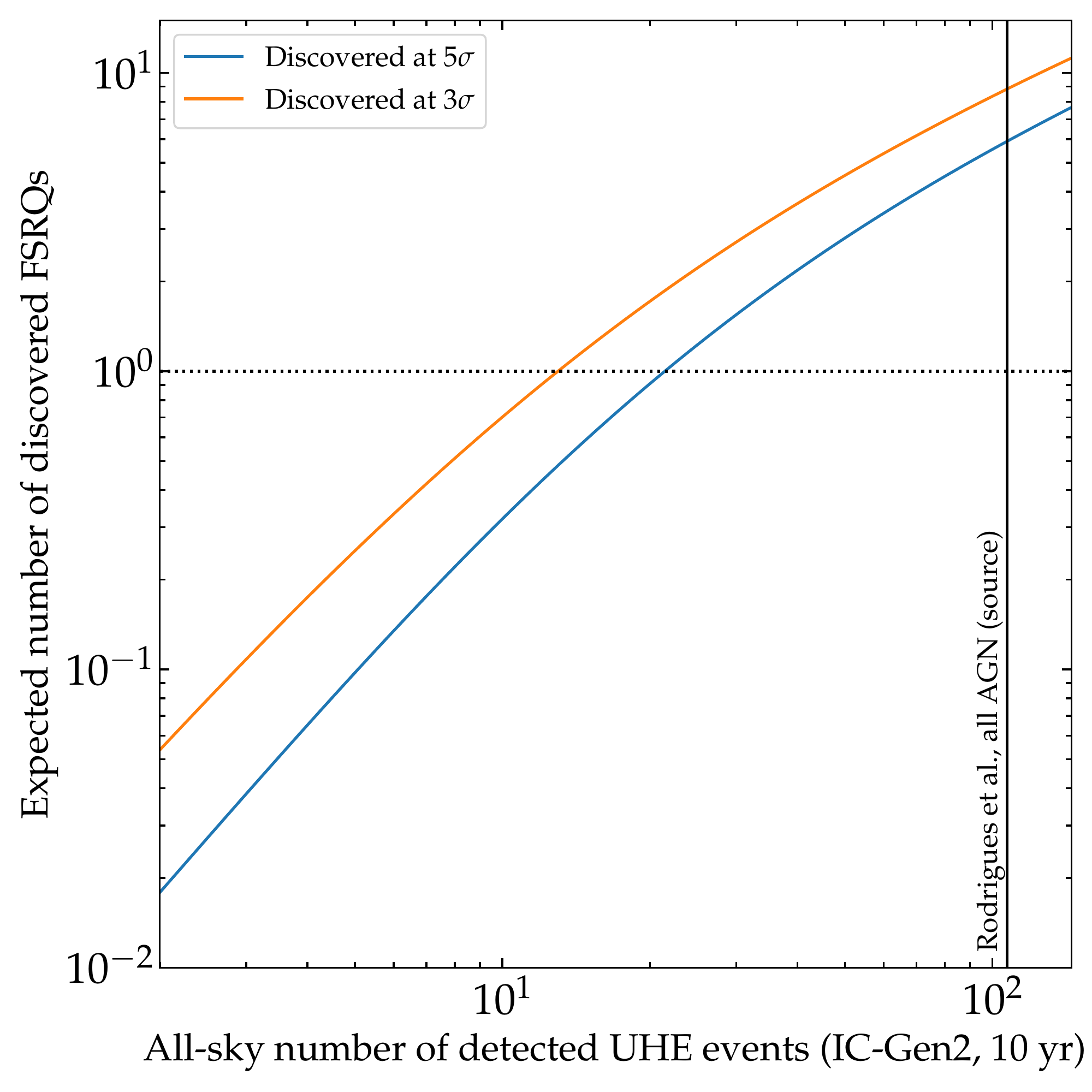}
 \caption{Expected number of FSRQ-like sources detected after 10 years of exposure in the IceCube-Gen2 radio array, as a function of the all-sky number of detected neutrino events. The neutrino luminosity is assumed to scale with the gamma-ray luminosity following \Refe~\cite{Rodrigues:2017fmu}. We adopt the luminosity and redshift evolution of the source number density from \Refe~\cite{Ajello:2015mfa}.  For reference, we mark the threshold value of one detected source and the expected all-sky number of events for the blazar model of \Refe~\cite{Rodrigues:2020pli}.  We show results  for $3\sigma$ and $5\sigma$ detection.}
 \label{fig:expecteddetectionfsrqs}
\end{figure*}

The results shown in the main text are based on the assumption of sources having all the same luminosity in their rest frame, and evolving in redshift according to the star formation rate. While this may be a reasonable assumption for some of the sources we consider, it does not apply to some of the most promising candidates, including BL Lacs and TDEs, which have a negative evolution with redshift, and FSRQs, which have a significant luminosity evolution with redshift. Here we apply our methods to study how the results change by relaxing our assumptions on the evolution.

First, we focus on jetted TDEs.  We extract their redshift evolution from Fig.~14 of \Refe~\cite{Kochanek:2016zzg}, for a black hole mass of $10^6$--$10^7$ solar masses. The projected constraints for this redshift evolution are shown for transient sources in \figu{pointsourcesensitivityjettedtde}.  Compared to \figu{pointsourcesensitivity}, which was generated assuming SFR redshift evolution, in \figu{pointsourcesensitivityjettedtde} we find that jetted TDEs still lie in the regions where they could be excluded by the absence of multiplets, provided they saturate the diffuse flux for our intermediate diffuse background model.  The question of whether TDEs are able to produce a diffuse flux comparable to the intermediate diffuse background  model depends on their neutrino luminosity, depends on specifics of the neutrino production model~\cite{Lunardini:2016xwi, Biehl:2017hnb, Winter:2020ptf, Reusch:2021ztx, Winter:2022fpf}; we do not investigate this question here.

Next, we turn to the case of blazars. We focus on FSRQs only, since, as discussed in Refs.~\cite{Rodrigues:2017fmu,Rodrigues:2020pli}, the UHE neutrino production from blazars is expected to be dominated by them (see also Refs.~\cite{Atoyan:2001ey,Atoyan:2002gu,Palladino:2018lov,Righi:2020ufi}). For a detailed treatment of FSRQs, a crucial element is the evolution with redshift of the luminosity distribution.  Therefore, we change our approach compared to the main text in the following way: in place of \equ{redshiftdistribution}, we use a correlated redshift and luminosity probability distribution extracted from Ref.~\cite{Ajello:2015mfa}. For this calculation, the local source density is not a free parameter that we let float, but is determined by the distribution of blazars detected by {\it Fermi}-LAT. Therefore, throughout the rest of the calculation, all the averages over the redshift distribution become averages over the redshift {\it and} luminosity distribution. The distributions of \Refe~\cite{Ajello:2015mfa} are in terms of the gamma-ray luminosity of FSRQs, whereas for the determination of the point source sensitivity we need the neutrino luminosity. We assume that the two are related by an efficiency $\epsilon_\nu$, \ie, $L_\nu=\epsilon_\nu(L_\gamma)L_\gamma$.
We assume that the efficiency depends on the gamma-ray luminosity following the dependence found in \Refe~\cite{Rodrigues:2017fmu}, Fig.~15, for proton injection and advective escape. However, we let its normalization float. Finally, following the procedure described in the main text, we obtain the expected number of sources discovered in 10 years. This number depends on the normalization of the neutrino production efficiency: a higher efficiency leads to a larger diffuse neutrino flux, more events detected and, therefore, more sources discovered.

Figure~\ref{fig:expecteddetectionfsrqs} shows the expected number of detected sources as a function of the all-sky expected number of events.  In this case, a point source can be discovered even if there are only 10--20 over the full sky, since the  sources are especially bright.  For the neutrino production model of \Refe~\cite{Rodrigues:2020pli}, which predicts about 108 events over all sky, a sizable number of point source discoveries is expected.


\bibliographystyle{JHEP.bst}
\bibliography{references.bib}

\end{document}